\documentclass[]{emulateapj}

\def\hubunits{{\rm km}\,{\rm s}^{-1}\,{\rm Mpc}^{-1}}
\def\PNM{P(N|M)}
\def\PNN{P(N|\langle N\rangle)}
\def\sigM{\sigma_{\log M}}

\def\dlgc0{{\Delta\log c_0}}
\def\meanN{\langle N \rangle}

\def\ngavg{\bar{n}_g}
\def\Navg{\langle N(M) \rangle}
\def\Nsat{\langle N_{\rm sat}(M) \rangle}
\def\Ncen{\langle N_{\rm cen}(M) \rangle}
\def\NsatNsatm1{\langle N_{\rm sat}(M)[ N_{\rm sat}(M) -1 ] \rangle}
\def\Mmin{M_{\rm min}}
\def\NNm1{\langle N(M)[N(M)-1] \rangle}
\def\Rvir{R_{\rm vir}}
\def\intdn{\int_0^\infty dM\frac{dn}{dM}}
\def\ngrp{n_{\rm grp}}
\def\Mviravg{\langle M_{\rm vir}(N) \rangle}
\def\hMsun{h^{-1}M_\odot}
\def\hMpc{h^{-1}{\rm Mpc}}
\def\hmpc{h^{-1}\,{\rm Mpc}}

\def\Pgghh{P_{\rm gg}^{\rm 2h}}

\def\k{{\bf k}}

\def\intdna{\int_0^\infty dM_1\frac{dn}{dM_1}}
\def\intdnb{\int_0^\infty dM_2\frac{dn}{dM_2}}

\def\xigg{\xi_{\rm gg}}
\def\ximm{\xi_{\rm mm}}
\def\xigm{\xi_{\rm gm}}
\def\xis{\xi_{\rm gg}^{\rm 1h}}
\def\xid{\xi_{\rm gg}^{\rm 2h}}

\def\xigmh{\xi_{\rm gm}^{\rm 1h}}

\def\intdn{\int_0^\infty dM\frac{dn}{dM}}

\def\dchi2{{\Delta\chi^2}}

\slugcomment{Received 2005 December 2; accepted 2006 December 23; ApJ 659}

\begin{document}

\shortauthors{Zheng \& Weinberg}
\shorttitle{Breaking the Degeneracies Between Cosmology and Galaxy Bias}

\title{Breaking the Degeneracies Between Cosmology and Galaxy Bias}
\author{
\textsc{Zheng Zheng}\altaffilmark{1,2} \textsc{and}
\textsc{David H. Weinberg}\altaffilmark{3}
}
\altaffiltext{1}{Institute for Advanced Study, Princeton,
NJ 08540; zhengz@ias.edu.}
\altaffiltext{2}{Hubble Fellow.}
\altaffiltext{3}{Department of Astronomy, Ohio State University, Columbus, 
OH 43210; dhw@astronomy.ohio-state.edu.}

\begin{abstract}
Adopting the framework of the halo occupation distribution (HOD), we 
investigate the ability of galaxy clustering measurements to simultaneously 
constrain cosmological parameters and galaxy bias. Starting with a fiducial
cosmological model and galaxy HOD, we calculate spatial clustering 
observables on a range of length and mass scales, dynamical clustering 
observables that depend on galaxy peculiar velocities, and the galaxy-matter 
cross-correlation measurable by weak lensing. We then change one or more 
cosmological parameters and use $\chi^2$-minimization to find the galaxy 
HOD that best reproduces the original clustering. Our parameterization of 
the HOD incorporates a flexible relation between galaxy occupation numbers 
and halo mass and allows spatial and velocity bias of galaxies within dark 
matter halos. Despite this flexibility, we find that changes to the HOD 
cannot mask substantial changes to the matter density $\Omega_m$, the matter 
clustering amplitude $\sigma_8$, or the shape parameter $\Gamma$ of the 
linear matter power spectrum --- cosmology and bias are not degenerate. With 
the conservative assumption of 10\% fractional errors, the set of observables
considered here can provide $\sim 10\%$ ($1\sigma$) constraints on 
$\sigma_8$, $\Omega_m$, and $\Gamma$, using galaxy clustering data 
{\it alone}. The combination $\sigma_8\Omega_m^{0.75}$ is constrained to
$\sim 5\%$. In combination with traditional methods that focus on 
large-scale structure in the ``perturbative'' regime, HOD modeling can 
greatly amplify the cosmological power of galaxy redshift surveys by taking 
advantage of high-precision clustering measurements at small and 
intermediate scales (from sub-Mpc to $\sim 20\hmpc$). At the 
same time, the inferred constraints on the 
galaxy HOD provide valuable tests of galaxy formation theory.
\end{abstract}

\keywords {cosmology: theory -- dark matter -- galaxies: formation 
-- galaxies: halos -- large-scale structure of universe}

\section{Introduction}

From the 1970s through the early 1990s, studies of galaxy clustering drove
much of the progress in cosmology.  Measurements of steadily improving 
dynamic range and precision demonstrated good agreement with the predictions
of a cosmological model incorporating scale-invariant, Gaussian primeval
fluctuations modulated by the transfer function expected in a universe
dominated by cold dark matter (CDM) with $\Omega_m h \sim 0.2$
(where $\Omega_m$ is the matter density parameter and 
$h \equiv H_0/100\;\hubunits$).
The advent of multi-fiber galaxy redshift surveys and improved
photometric input catalogs has dramatically improved the precision
of clustering measurements over the last decade, beginning with
the Las Campanas Redshift Survey (LCRS; \citealt{Shectman96}) and
continuing with the Two-Degree Field Galaxy Redshift Survey
(2dFGRS; \citealt{Colless01}) and the Sloan Digital Sky Survey
(SDSS; \citealt{York00,Abazajian04}).  In parallel, numerical simulations
and numerically tested analytic approximations have turned the task of
calculating non-linear dark matter clustering from specified initial 
conditions into an essentially solved problem.  The principal obstacle
to drawing cosmological inferences from
galaxy clustering measurements is 
now the uncertainty in the relation between the distribution of observable
galaxies and the underlying distribution of dark matter, the problem
known as galaxy bias.  Much of the cosmological progress in the last 
decade has been driven by observations that circumvent this complication,
such as cosmic microwave background (CMB)
anisotropies, weak gravitational lensing, the Ly$\alpha$ forest,
and the Type Ia supernova diagram.

These new observables favor an inflationary, low-$\Omega_m$, CDM-dominated 
model similar to that originally suggested by
galaxy clustering, which in turn implies that the galaxies that dominate
typical optically selected galaxy surveys must be approximately unbiased,
in the sense that the rms galaxy count fluctuations are similar to the
rms dark matter density fluctuations on large scales (see, e.g.,
\citealt{Lahav02}).
However, observed galaxy clustering varies
systematically with galaxy luminosity, color, and spectral or morphological
type (\citealt{Norberg02,Zehavi05}, and numerous references therein),
and reproducing the observed galaxy correlation function in an inflationary
CDM model requires that the bias of the correlation function vary with
separation on scales below a few megaparsecs 
(\citealt{Jenkins98,Zehavi04}).  The advances in observational
cosmology over the last few years have also raised the stakes for galaxy
clustering studies.  We are no longer interested in, for example, 
distinguishing $\Omega_m \sim 0.3$ from $\Omega_m=1$; instead, we
want to constrain $\Omega_m$ at the few percent level to increase the
power of tests for the nature of dark energy.  Despite improvements in
semi-analytic and numerical modeling of galaxy formation, it is not
clear that these methods predict galaxy bias robustly enough for this
kind of precision cosmology.  Faced with these challenges, most cosmological
applications of the 2dFGRS and the SDSS have focused on the
linear or near-linear regime, where generic arguments suggest 
that the effects of galaxy bias should be relatively simple.
These ``perturbative'' analyses of large-scale structure play a
significant role in the current web of cosmological constraints
(e.g., \citealt{Percival02,Spergel03,Tegmark04b,Cole05,Seljak05a,Tegmark06}), 
but they
are restricted to large scales where even these enormous surveys
have limited statistical precision.

\begin{figure}
\epsscale{1.2}
\plotone{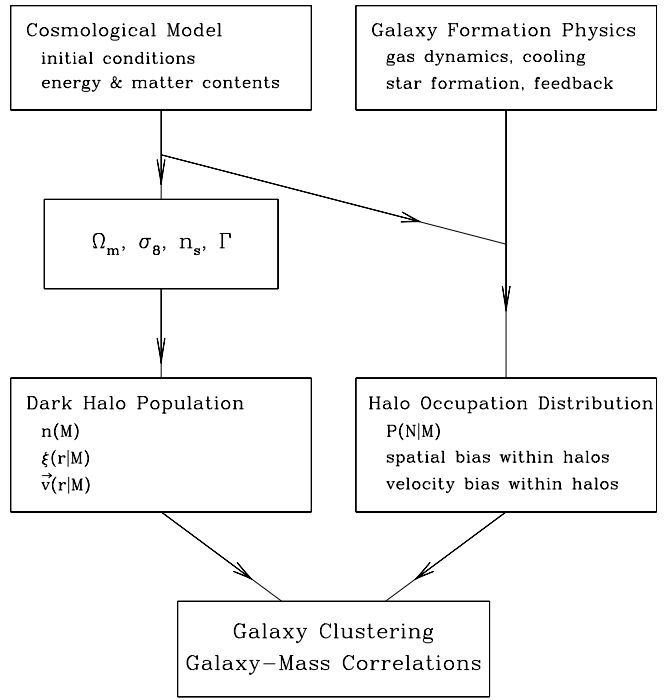}
\epsscale{1.0}
\caption[]{\label{fig:flowchart}
Interplay between the cosmological model and galaxy formation physics
in determining observable galaxy clustering.  The cosmological model
determines the mass function and clustering of the dark halo population.
Galaxy formation physics, operating within this cosmological model,
determines the HOD of different galaxy classes.
The clustering of any given class of galaxies can be predicted from
the halo population and the HOD.  In this paper we investigate how well one
can use observations of clustering and galaxy-mass correlations to
infer cosmological and HOD parameters simultaneously.
}
\end{figure}

In this paper we argue that recent developments in the theoretical 
description of galaxy bias allow a more aggressive approach to inferring
cosmological constraints from galaxy clustering measurements.
We work in the framework of the halo occupation distribution (HOD), which
characterizes galaxy bias in terms of the probability distribution
$P(N|M)$ that a dark 
matter halo of virial mass $M$ contains $N$ galaxies of a specified type,
together with prescriptions for the spatial and velocity bias of 
galaxies within dark matter halos 
\citep{Ma00,Peacock00,Seljak00,Scoccimarro01,Berlind02,Cooray02}.
Here the term ``halo'' refers to a bound dark matter structure of typical
overdensity $\rho/\bar{\rho} \sim 200$, in approximate dynamical equilibrium,
which may be the individual halo of a single bright galaxy or the common
halo of a galaxy group or cluster.\footnote{We have in mind the kinds of 
structures identified in $N$-body simulations by a friends-of-friends algorithm
with linking length $l \sim 0.15-0.2\bar{n}^{-1/3}$, but the precise
definition of halo does not matter provided that one is consistent
throughout all calculations.}  
The flow diagram in Figure~\ref{fig:flowchart}, adapted from 
\cite{Weinberg02}, sketches the interplay between the ``cosmological model''
and the ``physics of galaxy formation'' in determining observable galaxy
clustering, which we take to include both the traditional statistics 
measured from redshift surveys and the galaxy-matter correlations 
measured by galaxy-galaxy lensing 
(e.g., \citealt{Fischer00,Hoekstra01,Sheldon04}).
On the left side, the cosmological model, which specifies the 
initial conditions and the energy and matter contents of the universe, 
determines the mass function, spatial correlations,
and velocity correlations of the dark halo population.  
The intervening box indicates that the only features of the cosmological 
model that really matter in this context are $\Omega_m$ and the amplitude 
and shape of the linear matter power spectrum $P(k)$, here represented
by $\sigma_8$ (the rms linear matter fluctuation in $8\hmpc$ spheres),
the inflationary spectral index $n_s$, and the transfer function shape
parameter $\Gamma$, which itself depends on the values
of $\Omega_m$, $h$, and the baryon density \citep{Bardeen86,Hu96}.
Other features of the cosmological
model, such as the energy density and equation of state of the vacuum
component, may have an important impact on other cosmological
observables or on the
{\it history} of matter clustering, but they have virtually no effect on
the halo population at $z=0$, if the shape of $P(k)$ and the present day 
value of $\sigma_8$ are held fixed (see \citealt{Zheng02}).

On the right side of Figure~\ref{fig:flowchart}, the box titled
``galaxy formation physics'' represents the additional processes ---
such as shock heating, radiative cooling, star formation, feedback, and 
mergers --- that are essential to producing galaxies and determining
their masses, luminosities, diameters, colors, and morphologies.
These physical processes operate in the background provided by the 
evolving halo population, so together with the cosmological model they
determine the HODs of different classes of galaxies.
The halo population and the HOD together determine galaxy clustering
and galaxy-mass correlations.  One can, of course, use hydrodynamic
simulations or semi-analytic models to predict galaxy clustering statistics
directly, without computing the HOD as an intermediate step 
(e.g., \citealt{Kauffmann99,Benson00,Cen00,Pearce01,Weinberg04}).
However, these predictions reflect the combination of the cosmological
model and the galaxy formation theory, and unless one has complete 
confidence in the latter, one cannot draw secure conclusions about the former.
We would therefore like to know how well cosmological parameters can
be constrained {\it without} relying on a detailed theory of galaxy
formation, by using the data themselves to determine the relation between
galaxies and dark matter.

For this purpose, the HOD formulation of bias has two key strengths,
both emphasized by \cite{Berlind02}.  The first is the division of
labor implied by Figure~\ref{fig:flowchart}: galaxy formation physics
influences the HOD, but the properties of the halo population
defined at $\rho/\bar{\rho} \sim 200$ are determined almost entirely
by the much simpler physics of gravitational clustering, which can be
modeled accurately using $N$-body simulations or numerically tested 
analytic approximations.  The second is completeness: given a cosmological 
model and a specified HOD, one can calculate any galaxy clustering 
statistic on any scale, by populating $N$-body halos or by drawing on a 
steadily expanding array of analytic techniques.  The HODs of different 
galaxy classes therefore encode, statistically, all the aspects of galaxy 
formation physics that are relevant to predictions of galaxy clustering.  
(We discuss an important caveat to this statement below.) HOD modeling 
enables one to take full advantage of galaxy clustering data, with no 
restriction to large scales or particular statistics.

The goal of our empirical approach is to reverse the causal arrows
in Figure~\ref{fig:flowchart}, working backwards from the data to
properties of the HOD and the halo population, and from there to
conclusions about galaxy formation physics and fundamental aspects
of cosmology.
\cite{Berlind02} considered a fixed cosmological model and showed how 
changes to the HOD affect many of the traditional statistical measures
of galaxy clustering, such as correlation functions, the group multiplicity
function, and redshift-space distortions.  They argued that the 
complementary information from different statistics and different spatial
scales should permit an accurate empirical reconstruction of the HOD
for a specified cosmology, yielding physically informative tests of
galaxy formation theories.  \cite{Zheng02} showed that $\Omega_m$,
$\sigma_8$, and the power spectrum shape have non-degenerate effects
on the {\it halo} population, and they argued that changes to the galaxy HOD
could not fully mask the effects of cosmological parameter changes.
Here we complete this theoretical program by showing how galaxy
clustering measurements can constrain simultaneous changes to the
cosmological model and the galaxy HOD.  Our results add quantitative
teeth to the qualitative arguments and speculations in \cite{Berlind02}, 
\cite{Zheng02}, and \cite{Weinberg02}.

We define our cosmological parameter space by the values of $\Omega_m$,
$\sigma_8$, and $\Gamma$.  We concentrate mainly on $\Omega_m$ and $\sigma_8$
because $\Gamma$ (or, more generally, the shape of the linear power spectrum)
can be well constrained by combining the large-scale galaxy power
spectrum with other observables like CMB anisotropy and the Ly$\alpha$ forest.
We first compute predicted values of a number of galaxy clustering observables,
assuming a central model with observationally motivated choices of
cosmological parameters and the galaxy HOD.  We then change one or more
cosmological parameters, and we test how well this new cosmological model
can reproduce the original clustering ``data'' given complete freedom
to vary the HOD within a very flexible parameterization.  
To gain insight, we first explore several interesting axes within
the $(\Omega_m,\sigma_8,\Gamma)$ parameter space, varying parameters 
individually or in physically motivated combinations.
We arrive at
definite numbers by assuming that each of our 30 observables, some of
which represent the same clustering statistic measured at multiple
scales, can be measured with 10\% uncorrelated fractional uncertainty 
from a survey like the SDSS.  This assumption seems roughly plausible, 
but the ultimate strength of cosmological conclusions will
depend on the precision and dynamic range of the measurements.
Forecasting this precision for a survey like the SDSS, including the 
covariance of errors among different observables, is itself a major 
theoretical task, which we will not undertake here. 
For our theoretical investigation, we further restrict ourselves to observables
for which we have reasonable analytic approximations, and we suspect
that our quantitative conclusions will in the end prove overly pessimistic
because we must omit some observables that contain
significant additional information.

As discussed by \cite{Berlind02}, the completeness of the HOD as a 
description of bias rests on the assumption that the galaxy content of
a halo of virial mass $M$ is statistically independent of the halo's
larger scale environment.
\cite{Berlind03} show that this assumption accurately describes the galaxy
population in Weinberg et al.'s (\citeyear{Weinberg04}) hydrodynamic
cosmological simulation. It has fundamental theoretical roots in
the excursion set model of \cite{Bond91}, which predicts that the statistical
features of a halo's assembly history depend only on its present mass;
indeed, all semi-analytic galaxy formation models that use statistically
generated merger trees make this assumption implicitly.  Since the halo 
mass function itself varies with environment, with high-mass halos absent 
in low-density regions, models that tie galaxy populations
to halo masses still predict strong correlations between galaxy properties
and the large-scale environment, in good agreement with observations
(e.g., \citealt{Benson01,Berlind05,Zehavi05}).
However, while \cite{Lemson99} show that $N$-body halos of mass
$M \ga 10^{13}M_\odot$ have similar properties and formation 
histories in different environments, recent studies show a 
substantial correlation of halo formation redshift with large-scale 
overdensity for lower mass halos ($M \la 10^{12.5} M_\odot$),
contradicting the simplest form of the \cite{Bond91} model
(\citealt{Gao05,Harker06}; see also \citealt{Sheth04}).
As discussed below in \S\ref{sec:environment}, we expect the impact of
such dependence on the clustering statistics of mass- or 
luminosity-thresholded galaxy samples to be small, but given the level of 
precision we eventually hope to reach, it will probably be necessary
to allow for it when fitting the data.  Strategies for incorporating
environmental variations into HOD modeling will require guidance
from the next generation of hydrodynamic and semi-analytic
models of galaxy clustering, and we reserve this task for
future investigation.  Ultimately, the assumption of an
environment-independent HOD, or of any particular parameterization
of environmental dependence, must be tested empirically
by checking that the adopted model can consistently explain the
variations of galaxy clustering with large-scale environment.

Many current efforts to constrain cosmological parameters with galaxy 
clustering data make (explicitly or implicitly)
the much stronger assumption that galaxy bias can be 
adequately described by a linear model, $\delta_g=b\delta_m$, 
on the relevant scales.  Here $\delta_g$ and $\delta_m$ are the galaxy
and dark matter density contrast, respectively, and the linear bias
factor $b$ may depend on galaxy type but is assumed to be independent of scale.
In this case, $P_g(k)=b^2 P_m(k)$, and the shape of the galaxy power
spectrum provides powerful cosmological constraints in combination with
CMB data even if $b$ is unknown \citep{Percival02,Tegmark04b,Tegmark06}.
Large-scale redshift-space distortions measure 
$\beta \equiv \Omega_m^{0.6}/b$ \citep{Kaiser87,Hamilton98,Hawkins03},
which in combination with clustering measurements constrains
$\sigma_8\Omega_m^{0.6} = \Omega_m^{0.6}\sigma_{8,g}/b$, where 
$\sigma_{8,g}$ is the measured rms fluctuation of the {\it galaxy}
density contrast at $8\hmpc$.  The mass function or X-ray temperature
function of galaxy clusters constrains a similar combination of 
$\sigma_8$ and $\Omega_m$ \citep{White93} by a route that is independent
of galaxy bias.  At low redshift, the combination of galaxy-galaxy
lensing with the galaxy-galaxy correlation function constrains a
somewhat different combination of these parameters,
$\Omega_m/b \propto \sigma_8\Omega_m$, assuming a linear bias model
with $\xigg(r) = b\xigm(r) = b^2\ximm(r)$.
The triangle shape dependence of the reduced galaxy bispectrum can
constrain $b$ directly \citep{Fry94,Verde02}, in this case assuming
a quadratic bias model $\delta_g = b\delta_m + b_2\delta_m^2 + {\rm const.}$
to relate the galaxy and mass density fields at second order
\citep{Fry94,Juszkiewicz95}.

General theoretical arguments suggest that linear bias should be a good
approximation for the power spectrum on sufficiently large scales,
provided that the efficiency of galaxy formation is determined by
the local ($r < $ few Mpc) environment
\citep{Coles93,Fry93,Weinberg95,Mann98,Scherrer98,Narayanan00}.
Numerical experiments using such ``local'' bias models also show that
the ``$b$'' multiplying the power spectrum amplitude should be similar
to the ``$b$'' affecting redshift-space distortions \citep{Berlind01}.
Thus, the combination of these perturbative galaxy clustering
analyses can in principle yield separate constraints on $\Gamma$
(measured directly from the power spectrum shape), $\Omega_m$, 
and $\sigma_8$, with the degeneracy of $\Omega_m$ and $\sigma_8$
broken either by the bispectrum analysis or by the different
$\Omega_m$ dependence of the galaxy-galaxy lensing constraint and the
$\beta$ or cluster normalization constraints.  The applications of
these techniques to 2dFGRS and SDSS data show an impressive degree
of internal consistency and good agreement with external cosmological
constraints (\citealt{Percival02,Verde02,Spergel03,Tegmark04b,Cole05,
Seljak05a,Tegmark06}).  However, reliance on the linear or quadratic bias 
approximations restricts these analyses to large scales,
and it is not clear that these approximations hold consistently at
the level of accuracy desired for further improvements (better than 
10\%, say),
since scatter in the relation between galaxy and mass densities can
have different effects on different statistics and at different
scales \citep{Pen98,Dekel99}.
The HOD approach to galaxy clustering analysis 
substitutes a much more general model of galaxy bias, and it makes
use of high-precision measurements from small and intermediate scales 
(from sub-Mpc to $\sim 20\hmpc$) in addition to the lower 
precision measurements in the perturbative regime.
Furthermore, regardless of the strength of the cosmological constraints,
the HOD parameters derived from the data themselves provide detailed tests 
of galaxy formation models.

The program of HOD-based interpretation of observed
galaxy clustering is, in fact, well underway.
\cite{Jing98a} and \cite{Jing02} used HOD-type bias models to interpret
the correlation functions and pairwise velocity dispersions measured
from the LCRS and the Point-Source Catalog Redshift Survey (PSCz; 
\citealt{Saunders00}).  \cite{Peacock00}, \cite{Marinoni02}, and
\cite{Kochanek03} used the group multiplicity function to constrain
the galaxy occupations of high-mass halos. 
\cite{Guzik02}, \cite{Seljak05a},
and \cite{Mandelbaum06}
applied HOD modeling to galaxy-galaxy lensing measurements
from the SDSS.  \cite{Zehavi04}, analyzing a volume-limited sample of
bright ($M_r<-21$) SDSS galaxies, showed that HOD models naturally explain
the observed deviation from a power-law correlation function.
\cite{Zehavi05} measured the luminosity and color dependence of the 
SDSS galaxy correlation function and used it to infer the luminosity
and color dependence of the galaxy HOD, finding results in good qualitative
agreement with theoretical predictions \citep{Zheng05}.
\cite{Magliocchetti03} used a similar approach to investigate the halo
occupations of early- and late-type galaxies in the 2dFGRS,
while \cite{Collister05} derived HODs of red and blue 2dFGRS galaxies
from the group catalog of \cite{Eke04}, testing the consistency 
of their result by comparing predicted and observed correlation 
functions.
\cite{Abazajian05} used the SDSS measurements for $M_r<-21$ galaxies in
conjunction with CMB anisotropy data to infer simultaneous constraints
on HOD and cosmological parameters.
\cite{Tinker05} used HOD modeling of Zehavi et al.'s (\citeyear{Zehavi05})
clustering measurements to predict cluster mass-to-light ($M/L$) ratios,
and they inferred constraints on $\sigma_8\Omega_m^{0.6}$ by comparing
to published $M/L$ measurements.
HOD models have been applied to the interpretation of high-redshift
clustering by \cite{Bullock02}, \cite{Moustakas02}, \cite{Yan03}, 
\cite{Zheng04}, \cite{Ouchi05}, \cite{Lee06}, \cite{Coil06}, 
and \cite{Cooray06}.
Finally, \cite{Bosch03a} have initiated a comprehensive program
similar to the one described here, based on the closely related
conditional luminosity function (CLF) formalism 
(see also \citealt{Bosch03b,Bosch04,Bosch05,Bosch06,Yang03,Yang04,Yang05,Mo04,
Wang04}).
We discuss the similarities and differences between the HOD and CLF
approaches in \S~\ref{sec:discussion}.

In the next section we define our class of cosmological models,
list the analytic approximations we use for properties of 
the halo population, and describe our flexible parameterization of the 
galaxy HOD. 
In \S~\ref{sec:observables} we describe the methods 
and approximations that we adopt for analytic calculation of galaxy
clustering observables.
Section~\ref{sec:results} presents our main results, showing the 
ability of complementary galaxy clustering measurements to
constrain the galaxy HOD and cosmological parameters simultaneously.
Section~\ref{sec:environment} discusses the issue of 
environmental variations of the HOD.
Section~\ref{sec:discussion} summarizes our findings and discusses
the overall prospects for application of this approach.
A reader familiar with the field who wants an overview of our main 
results can read the first paragraph of \S\ref{sec:cosmology}, skim 
\S~\ref{sec:hodpar} to understand our HOD parameterization, then skip to 
\S~\ref{sec:results}, paying particular attention to \S~\ref{sec:omegam},
\S~\ref{sec:general}, and \S~\ref{sec:influence} and 
Figures~\ref{fig:chi2_omega}, \ref{fig:chgerr}, and \ref{fig:matrix}.

\section{Cosmological Model and HOD Parameterization}
\label{sec:model}

\subsection{Cosmological Parameters and Halo Properties}
\label{sec:cosmology}

Throughout this paper we adopt spatially flat $\Lambda$CDM cosmological 
models with Gaussian initial density fluctuations. The cosmological
model is defined by the mass density parameter $\Omega_m$ and the mass
fluctuation power spectrum $P(k) \propto k^{n_s}T^2(k;\Gamma)$, where $k$ 
is the wavenumber, $n_s$ is the spectral index of the inflationary power 
spectrum, and $T(k;\Gamma)$ is the transfer function with shape parameter 
$\Gamma$. We parameterize each cosmological model in this paper by
$\Omega_m$, $\sigma_8$, $n_s$, and $\Gamma$, where $\sigma_8$ is the rms
fluctuation of the linear density field filtered with a top-hat filter of
radius $8\hMpc$. We adopt the parameterization of \citet{Efstathiou92} for
the transfer function $T(k;\Gamma)$, which approximates the evolution of 
adiabatic primordial fluctuations. This formulation 
suffices for our purpose of investigating how galaxy clustering data
constrain HOD and cosmological parameters, with a single parameter $\Gamma$ 
encoding the combined effects of the matter-radiation transition (at the
scale $\lambda \propto \Omega_m h$) and the suppression of fluctuation 
growth in baryon or massive neutrino components. Observational analyses
should use a more accurate transfer function, e.g., from CMBFAST 
\citep{Seljak96}, but this would not change the sensitivity to the shape 
and amplitude of $P(k)$. It is important to note that we treat $\Omega_m$
and $\Gamma$ as independently variable quantities; we do not automatically
change the power spectrum shape when we change $\Omega_m$. In the context
of $\Lambda$CDM, our approach implicitly assumes that a change in $\Omega_m$ 
is compensated by a change in $h$ or some other parameter to keep the 
power spectrum at its empirically constrained shape.

Our calculations of galaxy clustering statistics rely on analytic
descriptions of the density profiles, mass function, spatial clustering,
and velocity statistics of dark matter halos.  We draw on the extensive
literature that presents numerically tested analytic approximations
or numerically calibrated fitting formulae for these quantities.

The density profile of a dark matter halo of mass $M$ is assumed to have 
the Navarro-Frenk-White (NFW) form (\citealt{Navarro95,Navarro96,Navarro97}),
\begin{equation}
\label{eqn:NFW}
\rho_m(r,M)=\frac{\rho_s}{(r/r_s)(1+r/r_s)^2},
\end{equation}
where the characteristic radius $r_s$ is related to the virial radius 
$\Rvir$ of the halo through the concentration parameter $c=\Rvir/r_s$,
and
\begin{equation}
\rho_s=\rho_0 \frac{\delta_{\rm vir}}{3}\frac{c^3}{\ln(1+c)-c/(1+c)}.
\end{equation}
We define the virial radius by the condition $\delta_{\rm vir}= 200$,
independent of cosmology, where $\delta_{\rm vir}$ 
is the average mass density of the halo
within $\Rvir$ in units of the mean matter density $\rho_0$.
The concentration parameter $c$ is a function
of halo mass. Here we adopt the relation given by \citet{Bullock01}
after modifying it according to our constant-$\delta_{\rm vir}$ halo
definition,
\begin{equation}
\label{eqn:concentration}
c(M)=c_0\biggl(\frac{M}{M_*}\biggr)^{\beta_c},
\end{equation}
where $c_0=11$, $\beta_c= -0.13$, and $M_*$ is the nonlinear mass.

The mass function of halos can be expressed as 
\begin{equation}
\label{eqn:halomf}
\frac{dn(M)}{dM} dM=\frac{\rho_0}{M}f(\nu)\frac{d\ln \sigma(M)}{dM} dM,
\end{equation}
where $n(M)$ is the space density of halos with mass higher than $M$,  
$\sigma(M)$ is the rms fluctuation of the mass overdensity at a mass scale 
$M$, $\nu=\delta_c/ \sigma(M)$ with $\delta_c\approx 1.686$ is the threshold 
density contrast for collapse, and $f(\nu)$ is a dimensionless function. 
The mass $M_*$ in equation~(\ref{eqn:concentration}) is defined by
the condition $\sigma(M_*)=\delta_c$ (i.e., $\nu=1$).
The function $f(\nu)$ can be approximately derived 
using the Press-Schechter (\citeyear{Press74}) or excursion set 
\citep{Bond91} formalism, or it can be calibrated by fitting the 
results of $N$-body simulations (e.g., \citealt{Sheth99,Jenkins01}). 
Here we use the fitting formula of \citet{Sheth99}, 
\begin{equation}
\label{eqn:fnu}
f(\nu)=A\sqrt{\frac{2a}{\pi}}\biggl[1+(a\nu^2)^{-p}\biggr]\nu 
       \exp\biggl(-\frac{a\nu^2}{2}\biggr),
\end{equation}
where $A=0.3222$, $a=0.707$, and $p=0.3$. This formula is consistent with the
ellipsoidal collapse model (\citealt{Sheth01b}), and it provides an excellent 
match to results from simulations (see, e.g., \citealt{Reed03}).

At large scales, the halo two-point correlation function 
$\xi_{\rm hh}(r)$ is biased relative to the matter two-point correlation
function $\xi_{\rm mm}(r)$ by a mass-dependent factor $b_h^2(M)$.
We adopt Sheth et al.'s \citeyear{Sheth01b} formula for $b_h(M)$,
based on the ellipsoidal collapse model,
\begin{eqnarray}
\label{eqn:halobias}
b_h(M) & = & 1 + \frac{1}{\sqrt{a}\delta_c}\biggl[\sqrt{a}(a\nu^2)+
\sqrt{a}b(a\nu^2)^{1-c}\nonumber \\
       &   & -\frac{(a\nu^2)^c}{(a\nu^2)^c+b(1-c)(1-c/2)}\biggr] ,
\end{eqnarray}
where $a=0.707$, $b=0.5$, $c=0.6$, and $\nu$ and $\delta_c$ have the same 
meaning as in equation~(\ref{eqn:halomf}).
Our calculation of the galaxy 2-point correlation function, discussed in
\S~\ref{sec:spatial}, also incorporates scale dependence
of halo bias.
Recent numerical work \citep{Seljak04,Tinker05} suggests that the
bias factors given by equation~(\ref{eqn:halobias}) are systematically
too high.  Since we use the same approximations to calculate the 
``observations'' of our central model and the ``predictions'' of others,
we do not expect moderate errors in $b_h(M)$ to alter our conclusions.
However, comparisons to real observational data should use the 
most accurate available bias factors.

The rms three-dimensional (3D) space velocity of halos of mass $M$ can 
be approximated as
\begin{equation}
\label{eqn:halovdisp1}
\sigma_h(M)=H_0\Omega_m^{0.6}\sigma_{-1}\sqrt{1-\sigma_0^4/\sigma_1^2\sigma_{-1}^2},
\end{equation}
where
\begin{equation}
\sigma_j^2(M)=\frac{1}{2\pi^2}\int_0^\infty dk ~ k^{2+2j} P(k) W^2[k R(M)],
\end{equation}
$W[k R(M)]$ is the Fourier transform of the top-hat filter on mass scale $M$, 
and the smoothing radius $R$ is determined through $M=4\pi\rho_0R^3/3$
\citep{Sheth01a}.
The one-dimensional (1D) rms velocity is simply the 3D one divided 
by $\sqrt{3}$.
The rms 1D velocity difference of halos of masses $M_1$ and $M_2$ at a
separation $r$ is
\begin{equation}
\label{eqn:halovdisp2}
\sigma_{h,{\rm 1D}}^2(M_1,M_2|r)
=\sigma_{h,{\rm 1D}}^2(M_1)+\sigma_{h,{\rm 1D}}^2(M_2)+\Psi(M_1,M_2|r),
\end{equation}
where $\Psi(M_1,M_2|r)$ represents the correlation term.
In our calculations (see \S\ref{sec:dynamical}), we ignore the
correlation term, which is an acceptable simplification
\citep{Sheth01a}.
We assume that the 1D velocity dispersion of dark matter {\it within} a halo
of mass $M$ has the value $(GM/2\Rvir)^{1/2}$ expected for a singular
isothermal sphere, independent of radius.
Moderate departures from this assumption can be accommodated
by the velocity bias parameter $\alpha_v$, described below.

\subsection{HOD Parameterization}
\label{sec:hodpar}

The HOD of a specified class of galaxies is defined by the probability
distribution $P(N|M)$ that a halo of mass $M$ hosts $N$ galaxies,
together with
prescriptions for the spatial and velocity distributions of galaxies
within halos relative to those of the dark matter particles.
For our purpose in this paper, we want a flexible parameterization
of the HOD that could describe the relevant features of any physically
reasonable model.  We want the constraints on the HOD to be imposed by
the clustering measurements rather than by our parameterization, and
we want to give each cosmological model its best shot at reproducing
these measurements before rejecting it.

Conceptually, it is useful to separate $\PNM$ into a mean occupation
function $\Navg$, defined by $\Navg = \sum_N N\,\PNM$, and a
distribution $\PNN$ at fixed halo mass.  There have been numerous
investigations of the predictions of semi-analytic models, hydrodynamic
simulations, and high-resolution $N$-body simulations for $\PNM$
(e.g., 
\citealt{Kauffmann97,Benson00,Seljak00,Scoccimarro01,Sheth01a,White01,
Yoshikawa01,Cooray02,Guzik02,Scranton03}; \citealt{Berlind03,Kravtsov04,
Zheng05}; the last three are the most extensive and the most relevant to 
our considerations here).  
The predicted HOD depends on the defining characteristics
of the galaxy sample, such as luminosity, color, and morphological
type.  For the most sensitive cosmological investigations, it seems best
to work with volume-limited samples defined by luminosity thresholds,
which are the largest homogeneous samples (i.e., with the 
same distribution of galaxy types at all distances) 
that one can construct from an apparent magnitude-limited survey.
In this case, $\Navg$ should be a more or less monotonic function of
halo mass, since more massive halos contain, on average, either more galaxies
or brighter galaxies.  Below some cutoff mass, $\Navg$ should drop
to zero, as low-mass halos do not contain enough cooled baryons to make
a galaxy above the luminosity threshold.
Instead of a luminosity threshold, one can impose a threshold
in stellar mass, estimated for each galaxy from its luminosity
and its color or spectral energy distribution (e.g.,
\citealt{Bell01,Kauffmann03}).  This approach should reduce the
scatter between the galaxy observable and the host halo mass
by reducing the impact of stellar population variations, thus
sharpening the cutoff in $\Navg$ and minimizing its environmental
dependence.  However, if one starts from an
apparent magnitude-limited survey, then the mass-thresholded samples
one can create are smaller than the luminosity-thresholded samples
because of the range of galaxy colors.

While we do not want to impose strong theoretical priors on our HOD
parameterization, we have decided to incorporate one of the most
generic and useful results that comes from the galaxy formation 
papers cited above: the distinction between central and satellite
galaxies.  In hydrodynamic simulations and semi-analytic models,
most halos contain a galaxy near the center of mass, moving close
to the center-of-mass velocity, which is usually more massive and
older than any other galaxies in the halo \citep{Berlind03}.
\cite{Kravtsov04} show that the central-satellite distinction
naturally explains one of the quantitatively important properties
of the HOD, the sub-Poisson width of $\PNN$ at small mean occupation
numbers.  For satellite dark matter subhalos, the fluctuations about
the mean occupation are close to Poisson, but the number of central 
galaxies is by definition either zero or one, thus obeying
nearest-integer (Bernoulli) statistics.  Fluctuations in the
full $\PNN$ are substantially sub-Poisson as long as the central
galaxy makes a significant contribution to $\Navg$.
\cite{Zheng05} show that this argument carries over to galaxies
in hydrodynamic simulations and semi-analytic models.

Motivated by these results, we model the mean occupation function
for galaxies above a luminosity threshold as
\begin{equation}
\label{eqn:Navg}
\Navg = 
\frac{1}{2}\left[1+{\rm erf}
\left(\frac{\log M-\log\Mmin}{\sigM}\right)\right][1+S(M)],
\end{equation}
where the last set of square brackets represents the sum of one central 
galaxy and $S(M)$ satellites and the first provides a smooth cutoff
near $\Mmin$.
%\footnote{Throughout this paper, we use ``lg'' to denote
%a base-10 logarithm.}  
The form of the cutoff profile corresponds to a log-normal distribution
of central galaxy luminosity at fixed halo mass (see \citealt{Zheng05}).
We apply the same cutoff profile to satellite galaxy numbers, although
in practice these are already well below unity close to $\Mmin$.
To model $S(M)$ in a flexible way, we take the free parameters to
be the values of $\log S(M)$ at $m$ fixed values of $\log M$ and 
define the continuous function by a cubic spline that passes through
these $m$ values.  
We set $m=5$ and choose values of $\log S(M)$ at
$\log (M/\hMsun) = 11$, 12, 13, 14, and 15 for the spline fit.
Halos more massive than $10^{15}\hMsun$ are too rare to have much impact 
on the clustering statistics we consider here. Theoretical predictions of
$S(M)$ can be accurately described by a truncated power-law parameterized 
by a slope, amplitude, and low-mass cutoff (\citealt{Kravtsov04,Zheng05}),
but our five-parameter model allows for curvature or inflections that depart 
from these theoretical predictions.
In addition to the $S(M)$ values, the adjustable parameters of 
equation~(\ref{eqn:Navg}) are $\sigM$ and $\Mmin$.  We always assume
that the uncertainty in the galaxy mean density is negligible in 
comparison to the uncertainty in the clustering measurements.
Therefore, once we have specified $S(M)$ and $\sigM$, we choose the
value of $\Mmin$ that yields the correct mean number density, rather
than treating it as a free parameter.  (We could pick any of the
parameters for this treatment, but the mean density is most sensitive
to $\Mmin$.)

For the clustering measures and analytic approximations used in this 
paper, the only property of $\PNN$ that we need to specify is the
second factorial moment $\NNm1$, which determines the mean number
of pairs per halo.  The number of central galaxies is zero or
one with a relative probability determined by the mean $\Ncen$.
For satellite galaxies, we introduce a width parameter $\omega$
defined by
\begin{equation}
\label{eqn:NNm1sat}
\NsatNsatm1 = \omega \Nsat^2.
\end{equation}
The value $\omega=1$ corresponds to a Poisson distribution of
satellite numbers $N_{\rm sat}(M)$ at fixed halo mass, consistent with 
theoretical
predictions.  The parameter $\omega$ allows the satellite probability
to be broader or narrower than Poisson.  The full probability 
distribution becomes narrow at low $\Navg$ in any case because of
central galaxies.  The value of $\omega$ determines how quickly
the distribution approaches a Poisson-like width, thus influencing
the number of galaxy pairs in halos with mean occupation numbers
$\sim$ 1$-$several.  

In the hydrodynamic simulation examined by \cite{Berlind03}, satellite
galaxies have a radial profile similar to that of the dark matter.
However, the physical processes that determine the satellite profile
are complex (see, e.g., \citealt{Nagai05,Zentner05}), so there is no reason
to expect the profiles to be identical.  Here we assume that satellite
galaxies in a halo still follow an NFW profile, but we allow
the galaxy concentration parameter $c_g$ to differ from the
underlying dark matter concentration parameter $c$.  We characterize
the relative spatial distributions of satellite galaxies
and dark matter by $\dlgc0 \equiv \log c_g - \log c_0$, where a 
positive (negative) $\dlgc0$ means that galaxies are more (less)
centrally concentrated than dark matter.  We assume that central
galaxies always reside at the halo center of mass.

We also allow the velocity dispersion of satellites to differ from
the velocity dispersion of dark matter by an average factor $\alpha_v$,
which we assume to be independent of mass.  Central galaxies are
assumed to move at the halo center-of-mass velocity.  Simulations
suggest that $\alpha_v \approx 1$, but the results are not entirely
consistent on this point (e.g., \citealt{Berlind03,Kravtsov04}), and
in any event we want this dynamical question to be settled by
observations rather than by theory.  \cite{Tinker06} discuss the
influence of $\alpha_v$ on redshift-space distortions in detail.

Altogether we have 10 free parameters in our HOD prescription --- seven
in $\Navg$ ($\Mmin$, $\sigM$, five spline points),
one ($\omega$) for the second moment of $P(N|\Nsat)$,
one ($\dlgc0$) for the spatial bias within halos, and
one ($\alpha_v$) for the velocity bias within halos.
One of these (which in practice we take to be $\Mmin$) is determined
by matching the mean galaxy space density once the values of 
other parameters are specified.  The remaining nine are varied
to yield the best match to the observables described in 
\S\ref{sec:observables}.  The parameters $\omega$, $\dlgc0$,
and $\alpha_v$ could in principle depend on halo mass, but the variation 
would have to be quite strong over a fairly narrow mass range 
to have an effect that could not be adequately described by an
appropriate average value.

The main theoretical preconception that is built into our 
parameterization is the existence of central galaxies.
The central location of these galaxies makes a small but
not negligible difference to the two-point correlation
function (see \citealt{Berlind02}), although this could probably
be compensated by $\dlgc0$.  However, since $\Nsat$ in 
practice turns out to be small near the cutoff mass $\Mmin$,
the central galaxy parameterization necessarily leads to 
sub-Poisson $\PNN$ fluctuations at low $\Navg$.  Since this
aspect of galaxy formation physics is well grounded, we think
it is reasonable to adopt it, and it allows us to represent
a wider range of physically realistic HODs with our finite
parameter set.  However, even if we do not make the central-satellite
distinction and instead use a functional form like that of
\cite{Scranton03} to describe the width of $\PNN$ as a function
of halo mass, we reach similar conclusions about the ability
to break degeneracies between cosmology and galaxy bias.
In essence, the observations drive us to sub-Poisson fluctuations
at low halo masses whether or not we impose this trend as a
theoretical expectation (see \citealt{Benson00,Peacock00,Berlind02}).

\section{Analytic Calculation of Observable Quantities}
\label{sec:observables}

The observable quantities considered in this paper fall into three general 
categories. First are quantities related to the real-space clustering of 
galaxies: the rms galaxy number density contrast, the galaxy two-point 
correlation function, the reduced galaxy bispectrum, and the group 
multiplicity function. Second are dynamical quantities that depend on 
galaxy peculiar velocities: the large-scale redshift-space distortion 
parameter $\beta$, the galaxy pairwise velocity dispersion, and the 
estimated virial mass of galaxy groups as a function of group richness.
Finally, we include the galaxy-matter two-point cross-correlation function,
which can be measured by galaxy-galaxy lensing.  
Most of these quantities are continuous functions of scale or group
multiplicity.  To obtain a discrete set of observables, we sample
these functions at a discrete set of separations or multiplicities,
listed in \S\ref{sec:results}.  We space our samplings widely
so that we can treat the statistical errors in each measurement
as approximately uncorrelated.

In addition to these observables, we impose the requirement that 
the HOD model reproduce the correct value of the sample's mean space
density, implying
\begin{equation}
\ngavg=\intdn\Navg.
\end{equation}
In effect, we treat $\ngavg$ as an observable with negligible 
statistical error.

In this paper we limit ourselves to observables for which we have some
reasonable analytic approximation.  In many cases, these approximations
are not accurate enough for application to high-precision clustering
measurements from large surveys like the SDSS or 2dFGRS --- or at least
their accuracy has not been extensively tested with numerical simulations.
However, we use the same approximations to calculate ``observed''
quantities for our central cosmological model and HOD and ``predicted''
quantities for altered models.  Provided that our approximations
realistically capture the variations of the observables with variations
in the cosmological and HOD parameters, they are adequate for our
present purposes of forecasting the ability of clustering measurements
to break the degeneracy between cosmology and galaxy bias.
Observational applications of these techniques will require
further testing and development of analytic approximations.
They could also benefit from analytic approximations for other
statistics, or from numerical studies that provide accurate
interpolation formulae in the relevant ranges of cosmological and
HOD parameters.

\subsection{Spatial Clustering}
\label{sec:spatial}

At large scales, the galaxy two-point correlation function $\xigg(r)$
is biased relative to the matter correlation function by (the square of)
a number-weighted average of the halo bias factor,
\begin{equation}
\label{eqn:bg}
b_g=\frac{1}{\ngavg}\intdn \Navg b_h(M) ~.
\end{equation}
We use equations~(\ref{eqn:halomf}) and~(\ref{eqn:fnu}) for 
$dn/dM$ and equation~(\ref{eqn:halobias}) for $b_h(M)$.
The fractional rms fluctuation $\sigma_g(R)$ of galaxy number counts
in spheres of radius $R$ can be expressed as an integral over $\xigg(r)$,
so at large $R$ it is related to the rms mass fluctuation $\sigma(R)$ by
\begin{equation}
\label{eqn:sig_g}
\sigma_g(R)=b_g\sigma(R)~.
\end{equation}
At $R=8\hmpc$, the smaller of the two scales we use for $\sigma_g(R)$,
the approximation~(\ref{eqn:sig_g}) is good but not perfect
(see \citealt{Tinker06}).

More generally, $\xigg(r)$ can be decomposed into one-halo and two-halo terms,
$\xis(r)$ and $\xid(r)$, which respectively represent 
pairs of galaxies residing in the same halo and in separate halos.
The one-halo term dominates at small separations and the two-halo
term at large separations, with the transition occurring near the
virial diameter of large halos.
Noting that the total number of galaxy pairs [$\propto 1+\xigg(r)$] is 
simply the sum of the number of pairs from single halos 
[$\propto 1+\xis(r)$] and that from different halos [$\propto 1+\xid(r)$], 
we have  
$\xigg(r)=[1+\xis(r)]+\xid(r)$.

The one-halo term $\xis(r)$ can be exactly computed in real space through 
\citep{Berlind02}
\begin{eqnarray}
\label{eqn:xigg1h}
1+\xis(r) & = & \frac{1}{2\pi r^2\ngavg^2}
              \intdn\frac{\NNm1}{2} \nonumber \\
     &   & \times \frac{1}{2\Rvir(M)} F^\prime\left(\frac{r}{2\Rvir}\right),
\end{eqnarray}
where $\NNm1/2$ is the average number of pairs in a halo 
of mass $M$ and $F(r/2\Rvir)$ is the cumulative radial distribution of 
galaxy pairs, i.e., the average fraction of galaxy pairs in a halo of mass 
$M$ (virial radius $\Rvir$) that have separation less than $r$. The 
differential function 
$F^\prime(x)$ is determined by the profile of the galaxy distribution 
within the halo. We assume that there is always a galaxy located at the 
center of the halo, and others are regarded as satellite galaxies. 
With this assumption, $F^\prime(x)$ is the 
pair-number weighted average of the central-satellite pair distribution 
$F^\prime_{\rm cs}(x)$ and the satellite-satellite pair distribution 
$F^\prime_{\rm ss}(x)$ (see, e.g., \citealt{Berlind02,Yang03}),
\begin{eqnarray}
\label{eqn:weightedpair}
        & & \frac{\NNm1}{2} F^\prime(x) =
        \Nsat F^\prime_{\rm cs}(x) \nonumber \\
      & & +  \frac{\NsatNsatm1}{2} F^\prime_{\rm ss}(x).
\end{eqnarray}
The central-satellite galaxy pair distribution $F^\prime_{\rm cs}(x)$
is just the normalized radial distribution of galaxies within halos, 
and the  satellite-satellite galaxy pair distribution $F^\prime_{\rm ss}(x)$ 
can be derived through the convolution of the galaxy distribution profile 
with itself (\citealt{Sheth01a}). In the case of an NFW profile, as used in 
this paper, $F^\prime_{\rm ss}(x)$ can be analytically expressed 
(see Appendix~\ref{sec:appendixA}; also see \citealt{Sheth01a}).

At large scales, the two-halo term becomes $\xid(r) = b_g^2 \xi_{\rm mm}(r)$,
where $b_g$ is the galaxy number-weighted halo bias factor of
equation~(\ref{eqn:bg}).  At smaller scales, one must account for
the convolution of galaxy profiles over the finite size of halos,
scale dependence of the halo bias factor, and the fact that separate
halos by definition do not interpenetrate.  Because convolutions 
become multiplications in Fourier space, it is convenient to calculate
the two-halo component of the galaxy power spectrum $\Pgghh(k)$ and Fourier
transform it to get $\xid(r)$ \citep{Seljak00,Scoccimarro01}.
Our full method of calculating the two-halo term is described in
Appendix B of \cite{Tinker05}.  Our adopted treatment of halo exclusion
is referred to there as the ``spherical'' exclusion method, and
we use the \cite{Sheth01b} formula for $b_h(M)$ instead of the modified
formula proposed by \cite{Tinker05}.  We have used this
method in earlier papers \citep{Zehavi04,Zehavi05,Zheng04}.

Galaxies within the same dark matter halo constitute a galaxy group.
The cumulative multiplicity function of galaxy 
groups, which is the number density of galaxy groups that have $N$ or more 
members, is
\begin{equation}
\label{eqn:multi}
\ngrp(\ge N) = \sum_{i=N}^\infty \intdn P(i|M).
\end{equation}
In this work we make the simplifying approximation that $P(i|M)$ is
a nearest-integer distribution, where $N$ is 
one of the two integers bracketing $\Navg$, and the relative probability is
determined by having the right mean. 
The space density of galaxy groups with multiplicity
$\geq \Navg$ is approximately 
equal to the space density of halos with mass $\geq M$,
so given a cosmological model that specifies $n(M)$ one can effectively
``read off'' the high end of the mean occupation function
\citep{Peacock00,Berlind02,Marinoni02,Kochanek03}.
A full observational implementation should take into account both the
width of $\PNN$ and the scatter between estimated and true 
multiplicities introduced by the group finding algorithm.
We consider galaxy groups with
five or more members, so the approximation~(\ref{eqn:multi}) 
is fairly good.

\subsection{Dynamical Measurements}
\label{sec:dynamical}

In redshift space, the peculiar velocities of galaxies distort the
galaxy power spectrum and correlation function, producing anisotropy
in which the line of sight is a preferred direction.  By measuring
this distortion on large scales, where it is caused by coherent flows
into overdense regions and out from underdense regions, one can
infer a combination $\beta \equiv \Omega_m^{0.6}/b_g$ 
of the density parameter and the 
large-scale galaxy bias factor \citep{Kaiser87}. 
We adopt $\beta$ as the first of our dynamically sensitive observables. 
The task of inferring $\beta$ from clustering measurements in the
context of HOD models is discussed in detail by \cite{Tinker06}.

The next dynamical measurement we consider is the pairwise radial 
velocity dispersion of galaxies, defined as 
$\sigma_v^2(r) \equiv \langle v_{12}^2 \rangle - \langle v_{12} 
\rangle^2$, where $v_{12} \equiv - {\bf (v_1-v_2) \cdot (r_1-r_2) / 
|r_1-r_2|}$ is the radial (inward) velocity of a galaxy pair,
${\bf v}_i$ and ${\bf r}_i$ ($i=1,2$) are velocities and positions of 
the two galaxies, and the average is over all galaxy pairs with separations
near $r={\bf |r_1-r_2|}$.  The pairwise dispersion can be inferred
by modeling the redshift-space correlation function 
(e.g., \citealt{Bean83,Davis83}).

Similar to the galaxy correlation function, on small scales, $\sigma_v^2(r)$ 
is dominated by a one-halo term, which can be computed through 
\citep{Berlind02}
\begin{eqnarray}
\label{eqn:galvdisps}
\sigma_{v,{\rm 1h}}^2(r) & = & \frac{1}{2\pi r^2 \ngavg^2 \xigg(r)} 
                \intdn \frac{\NNm1}{2} \nonumber \\
                & & \times \frac{1}{2\Rvir} 
                F^\prime\left(\frac{r}{2\Rvir}\right) \sigma_{\rm gg}^2(M),
\end{eqnarray}
where $\sigma_{\rm gg}^2(M)$ should be understood as the pair-number weighted 
average dispersion of the relative radial velocities between pairs of
galaxies in the same halo.
If we assume that the velocity distribution of satellite galaxies 
is isotropic and isothermal and that the central galaxy is at rest with 
respect to the center-of-mass of the parent halo, then 
the pair-weighted average dispersion can be separated into
central-satellite and satellite-satellite terms as in 
equation~(\ref{eqn:weightedpair}),  
\begin{eqnarray}
\label{eqn:vpair}
& & \frac{\NNm1}{2} F^\prime \sigma_{\rm gg}^2(M) = 
\Nsat F_{\rm cs}^\prime \alpha_v^2\sigma_m^2(M) \nonumber \\
& & +
\frac{\NsatNsatm1}{2} F_{\rm ss}^\prime 2\alpha_v^2\sigma_m^2(M) ~.
\end{eqnarray}
Here $\sigma_m^2(M)=GM/2R_{\rm vir}$ is the 1D velocity dispersion of 
dark matter within the halo, $\alpha_v$ is the velocity bias factor, and other 
quantities have the same meaning as in equation~(\ref{eqn:weightedpair}).
Note that there is a factor of 2 for satellite-satellite velocity dispersions,
since both members of the pair are moving with respect to the halo
center-of-mass.  

On large scales, the two galaxies of each pair come from different halos,
so the relative motions of 
halos also contribute to the galaxy pairwise dispersion. We ignore the halo 
velocity correlation ($\Psi$ in eq.~[\ref{eqn:halovdisp2}]), which is an 
acceptable simplification (\citealt{Sheth01a}). Following the 
argument of \citet{Berlind02} but taking into account the 
separation of central and satellite galaxies, the galaxy pairwise radial 
velocity dispersion on large scales can be expressed as
\begin{eqnarray}
\label{eqn:galvdispl}
\sigma_{v,{\rm 2h}}^2(r) & = & \frac{1}{\ngavg^2[1+\xigg(r)]}
                      \intdna \nonumber \\
   & & \times \intdnb [1+\xi_{\rm hh}(r;M_1,M_2)] \nonumber \\
   & & \times [\langle N(M_1)\rangle \langle N(M_2)\rangle \sigma_{h,{\rm 1D}}^2(r;M_1,M_2) \nonumber \\
   & & \mbox{} +\langle N_{\rm sat}(M_1)\rangle \langle N(M_2)\rangle \alpha_v^2\sigma_m^2(M_1) \nonumber \\
   & & \mbox{} +\langle N(M_1)\rangle \langle N_{\rm sat}(M_2)\rangle \alpha_v^2\sigma_m^2(M_2)], 
\end{eqnarray}
where $\xi_{\rm hh}(r;M_1,M_2)$ is the two-point correlation function between
halos of mass $M_1$ and $M_2$ at separation $r$. The three terms inside the
last set of square brackets represent contributions by the pairwise
velocity dispersion 
of $M_1$ and $M_2$ halos, by the dispersion of 
satellite galaxies in $M_1$ halos, 
and by the dispersion of satellite galaxies in $M_2$ halos. 
The halo-halo velocity 
dispersion is imprinted on all $\langle N(M_1)\rangle \langle N(M_2)\rangle$
galaxy pairs. 
However, the velocity dispersion of satellite galaxies in, say, $M_1$ halos
only contributes to the dispersion of the 
$\langle N_{\rm sat}(M_1)\rangle \langle N(M_2)\rangle$ pairs involving
those satellites.
The halo-halo velocity dispersion can be computed through 
equation~(\ref{eqn:halovdisp2}), and 
on large scales we have $\xi_{\rm hh}(r;M_1,M_2)=b_h(M_1)b_h(M_2)\ximm$ 
and $\xigg(r)=b_g^2\ximm$.  With these substitutions,
equation~(\ref{eqn:galvdispl}) reduces to 
\begin{eqnarray}
\label{eqn:galvdisps2h}
\sigma_{v,{\rm 2h}}^2(r) & = & 
\frac{2}{1+\xigg(r)}
[
  \langle \sigma_{h,{\rm 1D}}^2\rangle_N 
 +\xigg(r)\langle \sigma_{h,{\rm 1D}}^2\rangle_{Nb}
] 
\nonumber \\
& & +
\frac{2\alpha_v^2}{1+\xigg(r)}
[
  \langle \sigma_m^2\rangle_{N_s} 
 +\xigg(r)\langle \sigma_m^2\rangle_{N_sb}
], 
\end{eqnarray}
where the number- and bias-weighted averages
$\langle\rangle_N$ and $\langle\rangle_{Nb}$ are defined as
\begin{equation}
\label{eqn:sig2N}
\langle f \rangle_N \equiv \frac{1}{\ngavg} 
                       \intdn \Navg f(M),
\end{equation}
\begin{equation}
\label{eqn:sig2Nb}
\langle f \rangle_{Nb} \equiv \frac{1}{\ngavg b_g}
                          \intdn \Navg b_h(M) f(M),
\end{equation}
and $\langle\rangle_{N_s}$ and $\langle\rangle_{N_sb}$ are defined
by replacing $\Navg$ with $\Nsat$ in equations~(\ref{eqn:sig2N}) and 
(\ref{eqn:sig2Nb}), respectively.
Since we do not have a good approximation for $\sigma_v^2(r)$ at
intermediate scales, we only consider pairwise dispersion measurements
at small scales, where equation~(\ref{eqn:galvdisps}) applies,
or on large scales, where equation~(\ref{eqn:galvdisps2h}) applies.
Ultimately, the constraints obtainable from anisotropies in
redshift space on small, intermediate, and large scales 
can be inferred using the methods of
\cite{Tinker06} and \cite{Tinker07}.

We also consider dynamical measurements of the average virial masses of galaxy 
groups of fixed multiplicity, $\Mviravg$.  These can be inferred from
galaxy positions and velocities using the kind of estimators described by
\cite{Heisler85}.  These estimators are affected by velocity bias, with
the estimated mass scaling as $\alpha_v^2$.  For a given halo mass function
and HOD, we therefore calculate this quantity as
\begin{equation}
\label{eqn:Mvir}
\Mviravg = \alpha_v^2 \frac{\int_0^\infty dM (dn/dM) P(N|M) M}{\int_0^\infty dM (dn/dM) P(N|M)}~.
\end{equation}
As in our multiplicity function calculation, we 
approximate $\PNN$ by a nearest-integer distribution.
With equation~(\ref{eqn:Mvir}), we implicitly assume that the mass 
estimators are applied only to satellite galaxies, or else that the
reduced dispersion due to the central galaxy is properly taken
into account.  With large enough galaxy numbers, more sophisticated
mass estimators can circumvent the effects of velocity bias
(e.g., \citealt{Carlberg97}).  In addition, average group masses
can be estimated using gravitational lensing or X-ray properties
instead of galaxy velocities.  We do not consider these kinds of 
$\alpha_v$-free mass estimates in our analysis, but it is clear that
they would add power to our overall set of constraints if they
could be implemented robustly.

\subsection{Galaxy-Mass Clustering}
\label{sec:xigm}

The galaxy-mass cross-correlation can be probed by galaxy-galaxy lensing
(e.g., \citealt{Fischer00,Hoekstra01,Sheldon04,Seljak05a}). 
In practice, what is usually inferred
from galaxy-galaxy lensing at low redshift 
is the product of the mean matter density 
$\Omega_m$ and the galaxy-mass two-point cross-correlation function 
$\xigm(r)$, so we adopt values of
$\Omega_m \xigm(r)$ as the observables in our analysis. 

The galaxy-mass two-point cross-correlation function can also be decomposed 
into one-halo and two-halo terms.  For simplicity, we only consider 
small-scale cross-correlation, where the one-halo term dominates. Similar to 
the one-halo term of the galaxy two-point correlation function, the one-halo 
term $\xigmh(r)$ of the galaxy-mass two-point cross-correlation function is 
computed in real space by counting galaxy and mass particle pairs,
\begin{eqnarray}
\label{eqn:xigm1h}
 & & 1+\xigmh(r) = \frac{1}{4\pi r^2\ngavg \Omega_m \rho_c} \nonumber \\
 & & \times  \intdn \Navg M
            \frac{1}{2\Rvir(M)} F_{\rm gm}^\prime\left(\frac{r}{2\Rvir}\right),
\end{eqnarray}
where $\rho_c$ is the critical density. The function $F_{\rm gm}(x)$ is the 
cumulative radial distribution of galaxy and mass particle pairs (not 
galaxy-galaxy pairs as in eq.~[\ref{eqn:xigg1h}]), which is determined by 
the distribution profiles of galaxies and dark matter within the halo. When 
the effect of the central galaxy is taken into account, we have (similar to 
eq.~[\ref{eqn:weightedpair}])
\begin{equation}
        \Navg F_{\rm gm}^\prime(x) =
        F_{\rm cm}^\prime(x)
      + \Nsat F_{\rm sm}^\prime(x), 
\end{equation}
where $F^\prime_{\rm cm}$ ($F_{\rm sm}^\prime$) is the cumulative 
distribution of pairs made of the central galaxy (satellite galaxies) and 
mass particles. The function $F_{\rm cm}^\prime(x)$ is simply the radial 
distribution of matter, which can be directly obtained from the NFW profile.
The function for the satellite-mass pairs, $F^\prime_{\rm sm}(x)$, can be 
derived by convolving satellite and dark matter profiles. Under our assumption 
that the satellite distribution inside a halo also follows an NFW profile, 
$F^\prime_{\rm sm}(x)$ is the convolution of two NFW profiles with 
different concentration parameters, which has an analytic expression
given in Appendix~\ref{sec:appendixA}.
More general discussions of galaxy-galaxy lensing in the halo
occupation framework are given by \cite{Mandelbaum05}, who 
focus on halo virial masses and satellite galaxy fractions,
and \cite{Yoo06}, who focus on the 
derivable constraints on $\Omega_m$ and $\sigma_8$.

\subsection{The Galaxy Bispectrum}
\label{sec:bispectrum}

At the final step of our analysis, we consider the large-scale 
behavior of the galaxy bispectrum, the Fourier transform of the 
three-point correlation function.  Although this is a measure of
real-space clustering, like those in \S\ref{sec:spatial}, we treat
it separately because its information content is rather different
and because it is more difficult to measure observationally.
The specific observable that we consider is the reduced bispectrum
\begin{equation}
Q(\k_1,\k_2,\k_3) = 
\frac{B(\k_1,\k_2,\k_3)}{P(k_1)P(k_2)+P(k_2)P(k_3)+P(k_3)P(k_1)} ~,
\end{equation}
where $B(\k_1,\k_2,\k_3)$ is the bispectrum, $P(k)$ is the power spectrum,
and the three wavevectors form a triangle ($\k_1+\k_2+\k_3=0$). In the 
weakly nonlinear regime, the local relation between the galaxy 
density contrast $\delta_g$ and the mass density contrast 
$\delta$ can be written as a Taylor expansion 
$\delta_g=\sum b_{g,n}\delta^n/n!$.  The 
reduced galaxy bispectrum $Q_g$ is related to the reduced matter bispectrum 
$Q_m$ through 
\begin{equation}
\label{eqn:qg}
Q_g(\k_1,\k_2,\k_3)=
\frac{Q_m(\k_1,\k_2,\k_3)}{b_{g,1}}+\frac{b_{g,2}}{b_{g,1}^2},
\end{equation}
which shows contributions from both gravitational clustering and galaxy 
bias (\citealt{Fry94}; see also \citealt{Fry93,Juszkiewicz95}).
On large scales, $b_{g,1}$ and $b_{g,2}$ are simply the galaxy 
number-weighted halo bias factors $b_{h,1}$ and $b_{h,2}$, respectively, where 
$b_{h,1}$ and $b_{h,2}$ are predicted by nonlinear perturbation theory 
\citep{Scoccimarro01}. The reduced matter bispectrum is calculated using the 
linear matter power spectrum $P(k)$ and the second-order perturbative 
bispectrum $B_m^{\rm PT}$ (\citealt{Fry84,Scoccimarro01}),
\begin{equation}
B_m^{\rm PT} = 2F_2(\k_1,\k_2)P(k_1)P(k_2)+{\rm cyc.},
\end{equation}
where $F_2(\k_1,\k_2)=\frac{5}{7}+
\frac{1}{2}\cos \theta_{\rm 12}(k_1/k_2+k_2/k_1)+ 
\frac{2}{7}\cos^2\theta_{\rm 12}$, with 
$\k_1\cdot\k_2=k_1 k_2\cos \theta_{\rm 12}$. The weak dependence on $\Omega_m$
(\citealt{Kamionkowski99}) is neglected here.
The key diagnostic power of $Q_g$ on large scales lies in the distinctive
triangle-shape dependence of $Q_m$ predicted by gravitational 
perturbation theory.  Positive linear bias ($b_{g,1} > 1)$ suppresses $Q_g$,
and non-linear bias ($b_{g,2}>0$) can restore the amplitude but not
the shape dependence \citep{Fry94,Verde02}.

\section{Breaking the Degeneracy Between Bias and Cosmology}
\label{sec:results}

Motivated by CMB anisotropy measurements (e.g., 
\citealt{Netterfield02,Pryke02,Spergel03}), the abundance of galaxy 
clusters (e.g., \citealt{White93,Eke96}), high-redshift supernova observations 
(e.g., \citealt{Riess98,Perlmutter99,Riess04}), and measurements of the 
3D galaxy power spectrum and the Ly$\alpha$ forest power spectrum
(e.g., \citealt{Croft02,Percival02,Tegmark04b,Seljak05b}), 
we assume 
a central cosmology with 
$\Omega_m=0.3$, $\sigma_8=0.9$, $n_s=1.0$, and $\Gamma=0.2$. 
We model a galaxy population of mean space density
$\ngavg=0.01 h^3{\rm Mpc}^{-3}$, similar to that of galaxies
with $r$-band absolute magnitude brighter 
than $-19.5+5\log h$ in the SDSS.  We therefore use  
Zehavi et al.'s (\citeyear{Zehavi05}) results for $M_r \leq -19.5 + 5\log h$
galaxies as a guide for our central model's HOD parameters. 
We assume a satellite occupation $S(M)=M/(1.4\times 10^{13}\hMsun)$
in equation~(\ref{eqn:Navg}), i.e., the number of satellites is 
proportional to the halo mass well above the cutoff $\Mmin$.
We set the central galaxy cutoff parameter $\sigM$ to 0.2, 
consistent with predictions from galaxy formation models (\citealt{Zheng05}). 
With these choices, the cutoff mass required to match 
$\ngavg=0.01 h^3{\rm Mpc}^{-3}$ is
$\Mmin=5.55\times 10^{11}\hMsun$. The probability distribution of satellite 
galaxies is taken to be a Poisson distribution, i.e., $\omega=1$ in 
equation~(\ref{eqn:NNm1sat}).  We assume that satellite galaxies 
have the same radial profile and velocity dispersion as dark matter
within halos, i.e., $\dlgc0=0$ and $\alpha_v=1$.

For the central cosmology and the central HOD, we use the analytic formulae 
presented in \S~\ref{sec:observables} to calculate a set of observables, which 
we thereafter treat as observational measurements.  These observables are
the rms galaxy overdensity $\sigma_g(r)$ on scales of 
$8 \hMpc$ and $15 \hMpc$; the group multiplicity function $\ngrp(\ge N)$ for 
$N=5,$ 10, 20, and 40; the galaxy two-point correlation function $\xigg(r)$
at $r=0.1,$ 0.3, 0.5, 1, 2, 5, and 10 $\hMpc$; the $\beta$ parameter; the 
galaxy pairwise velocity dispersion 
$\sigma_v(r)$ at $r=0.25,$ 0.5, 3, 5, and 10 
$\hMpc$; the average virial mass of galaxy groups $\Mviravg$ for $N=10,$ 20, 
40, and 80; the galaxy-mass cross-correlation function 
$\Omega_m\xigm(r)$ at $r=0.25$ and 0.5 $\hMpc$; and 
values of the reduced 
galaxy bispectrum on large scales, for wavevector triangles having two sides 
fixed at $k_2=2k_1=0.05 h {\rm Mpc}^{-1}$ and subtending angles
$\theta=0$, $\pi/4$, $\pi/2$, $3\pi/4$, and $\pi$. 
We assume that each of the 30 ``measurements'' has a $1\sigma$ fractional
uncertainty of 10\%, and that the measurement errors are uncorrelated.
On its own, the 10\% error assumption is probably pessimistic; even
the largest scale observables can probably be measured at least this
well with the full SDSS, and some of the smaller scale quantities have
already been measured more precisely (we investigate the effect of
reducing errors for selected observables in \S\ref{sec:general}).
 However, the assumption that
the measurement errors are uncorrelated is optimistic, even though we 
space the observables for any given statistic (e.g., multiple scales of 
the correlation function) fairly widely. Unless there are strong 
positive correlations among many observables, the constraints on the HOD 
and cosmology should not degrade much with respect to our forecasts, but
a full application to observations must include the covariances among 
observables, calculated from the data or from mock catalogs.

Each time we change cosmological parameters, and thus the statistics of the
halo population, we search for HOD parameters that yield the best possible
match to the observables of the central model.  Specifically, we minimize
\begin{equation}
\Delta \chi^2 = \sum_i 
   \frac{(F_{i}^{\rm p} - F_{i}^{\rm o})^2}{\sigma_{F_i}^2},
\end{equation}
where $F_{i}$ represents the $i{\rm th}$ observable with an observational error
$\sigma_{F_i}$, the superscript ``$p$'' indicates the predicted value
for a given cosmology and HOD model, and the superscript ``$o$'' 
indicates the observed value calculated for the central cosmological
model and HOD parameters.  To isolate the information content of 
different clustering statistics, we start by considering only the 
spatial clustering observables described in \S\ref{sec:spatial}, then
add new observables one at a time.  We use a Gauss-Newton scheme
to carry out the $\chi^2$ minimization
(see Appendix~\ref{sec:appendixB} for details). 
We characterize the acceptability of models in terms of $\Delta\chi^2$,
appropriate for ``flat'' priors that treat all parameter values equally.

\subsection{Constraining HOD Parameters for a Fixed Cosmology}
\label{sec:fixed}

Before turning to cosmological parameter constraints, we first ask how
well the HOD can be determined if we assume that the cosmological
parameters are known perfectly from independent data.  Observational
constraints on the HOD for an assumed cosmology have already been
derived from the 2dFGRS, the SDSS, and other data sets, but these
studies have all used restricted forms of the HOD and limited subsets
of the galaxy clustering observables, such as the correlation function
or the multiplicity function 
(e.g., \citealt{Jing98a,Peacock00,Marinoni02,Kochanek03,Lin04,Bosch03b,
Magliocchetti03,Zehavi04,Zehavi05,Collister05}).  
Here we adopt our much more flexible HOD parameterization and examine
the constraints that could be obtained with 10\% measurements of all
of the observables discussed above.  Like \citet{Yan03} and \citet{Bosch05}, 
we explore the HOD parameter space using a Monte Carlo Markov Chain 
(MCMC) technique (e.g., \citealt{Gilks96}). 
We adopt the central cosmological model throughout, and we start the chain
from the central HOD parameters, which by definition yield $\Delta\chi^2=0$.
At any point of the chain, we generate a 
new set of HOD parameters by taking a random walk in the HOD parameter    
space with the step size for each parameter drawn from a Gaussian 
distribution. We accept the new HOD model with a probability of 1 if 
$\chi^2_{\rm new}\leq\chi^2_{\rm old}$ and 
$\exp[-(\chi^2_{\rm new}-\chi^2_{\rm old})/2]$ if 
$\chi^2_{\rm new} > \chi^2_{\rm old}$, where $\chi^2_{\rm new}$ and 
$\chi^2_{\rm old}$ are values of $\chi^2$ for the new model and for the 
previous model, respectively. Flat priors are adopted for HOD 
parameters $S(M)$, $\sigM$ and $\omega$ in logarithmic space and 
$\dlgc0$ and $\alpha_v$ in linear space.

\begin{figure*}
\plotone{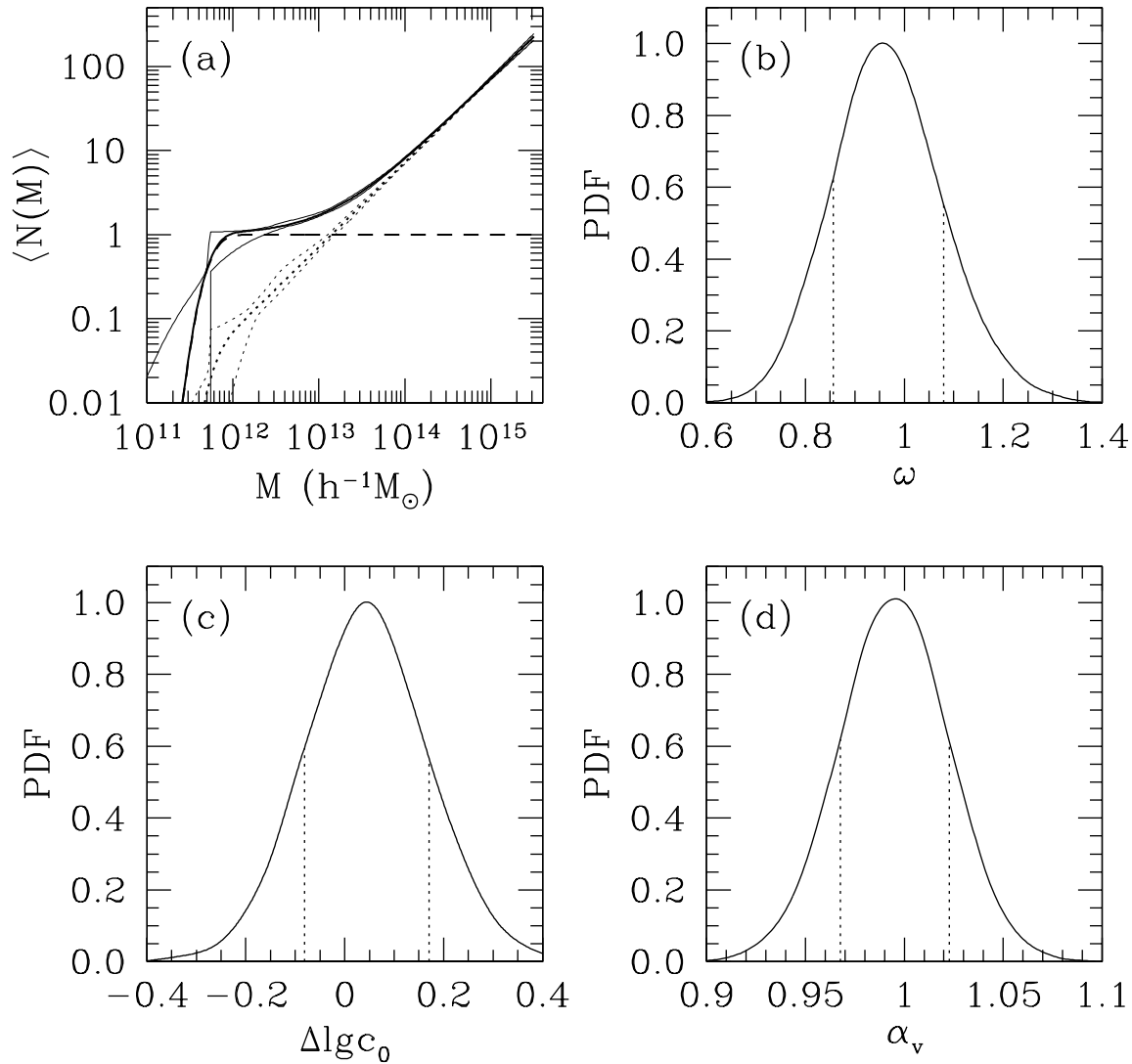}
\caption[]{\label{fig:mcmc_hodpar}
Uncertainties in the determination of the HOD with a known
cosmology. Panel ($a$) shows the mean occupation functions of
central galaxies ({\it dashed line}), satellite galaxies ({\it dotted
lines}), and all galaxies ({\it solid lines}). Thick lines are for the
input (central) HOD. Thin lines are envelopes of the mean occupation
functions determined from HOD models
with $\Delta\chi^2<1$, representing the 1-$\sigma$ range. The other three
panels show the marginalized likelihoods for the distribution width parameter
$\omega$, the halo concentration parameter $\Delta\log c_0$, and the
velocity bias $\alpha_v$, respectively. Vertical dotted lines in each of
these three panels mark the central 68.3\% of the distribution (i.e., the
1-$\sigma$ range).
}
\end{figure*}

The constraints on $\Navg$ and other HOD parameters are illustrated in 
Figure~\ref{fig:mcmc_hodpar}. 
Figure~\ref{fig:mcmc_hodpar}$a$ plots the envelope of 
$\Navg$ curves determined from HOD models that have $\Delta\chi^2<1$,
thus showing the 1-$\sigma$ uncertainty in $\Navg$. 
With our adopted set of observables, the shape of $\Navg$ near the cutoff 
mass is poorly constrained.  The mean galaxy density imposes a strong
constraint on the cutoff scale, but either hard or soft cutoffs can
produce the same $\ngavg$.  The halo bias factor is nearly independent
of mass in the low-mass regime
(see, e.g., \citealt{Jing98,Sheth01b,Seljak04}), and these halos contain
only a single galaxy, so variations in the cutoff profile do not
affect either the two-halo or one-halo terms of $\xigg(r)$.
Observables that we have not considered, like the void probability 
function or the Tully-Fisher relation, might provide stronger constraints
on the location and form of the $\Navg$ cutoff.

At higher halo masses, our observables provide much greater leverage,
and $\Navg$ is tightly constrained over the range $\meanN \sim 1.2$
to $\meanN \sim 100$.  The constraints loosen at still higher halo
masses, since these halos are too rare to have much impact on the
correlation function, and we only consider group virial masses up
to $N=80$ and the multiplicity function up to $N=40$.
Figures~\ref{fig:mcmc_hodpar}$b$ -- \ref{fig:mcmc_hodpar}$d$ 
show the marginalized likelihoods of the parameters 
$\omega$, $\Delta\log c_0$, and $\alpha_v$, with dotted vertical
lines marking the $1\sigma$ range.
Under our assumption of 10\% 
observational errors, the distribution width parameter $\omega$ can be 
constrained at the level of $10-15$\% (1-$\sigma$). For the difference 
between galaxy and dark matter concentrations, the 1-$\sigma$ 
constraint is $0.10-0.15$ dex, although with more small-scale points
in $\xigg(r)$ this quantity could be pinned down more tightly.
The 1-$\sigma$ uncertainty in the 
velocity bias $\alpha_v$ is only a few percent, 
largely because the spatial clustering observables tightly constrain
the true halo mass at fixed $N$, and the average virial masses
then determine the value of $\alpha_v$.

\begin{figure*}
\plotone{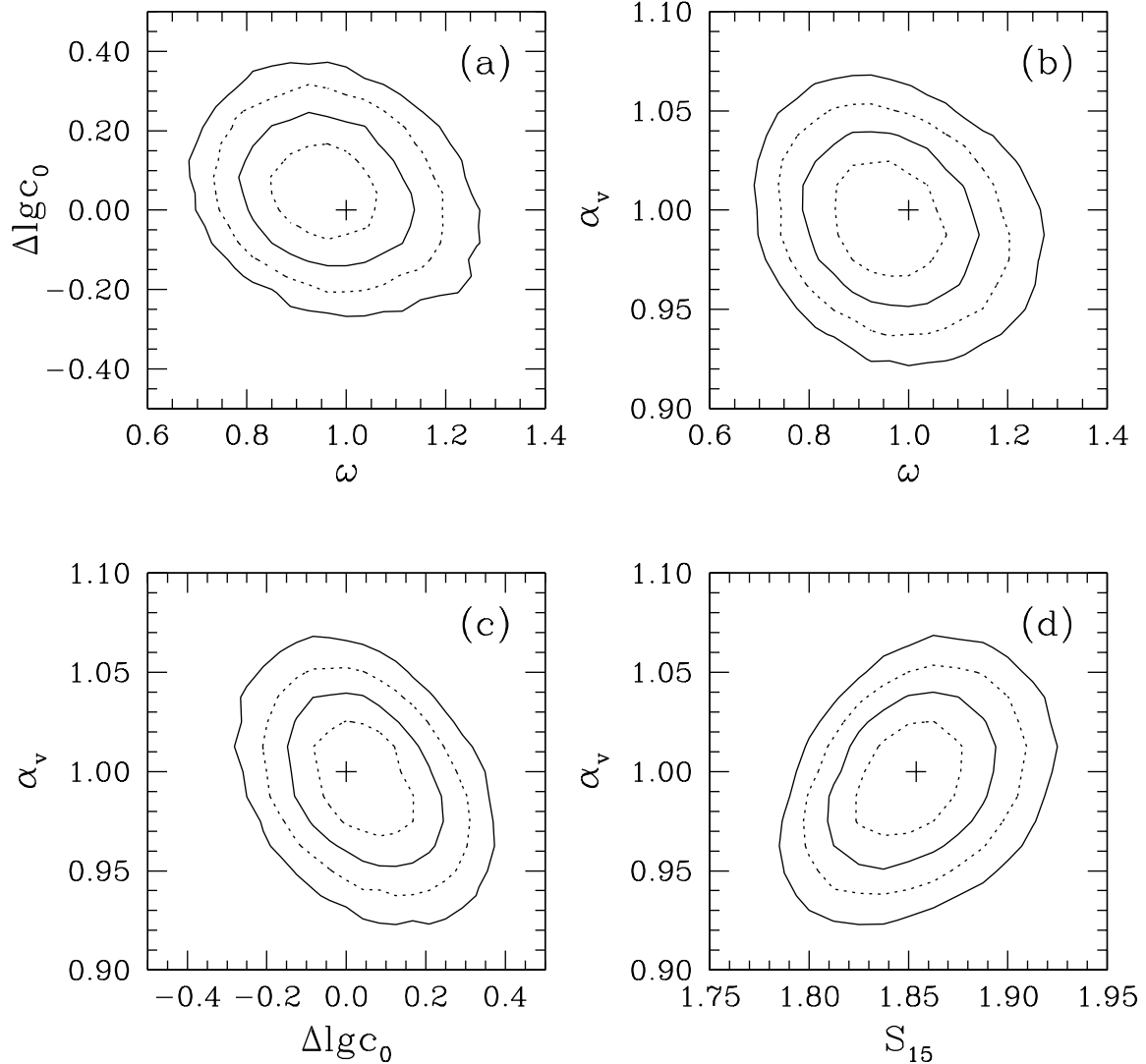}
\caption[]{\label{fig:mcmc_corr}
Marginalized contours and correlations between selected pairs of HOD
parameters for a known cosmology. The parameters plotted are the distribution
width parameter $\omega$, the halo concentration parameter $\Delta\log c_0$,
the velocity
bias $\alpha_v$, and the mean occupation number $S_{\rm 15}$ in halos of
mass $10^{15}\hMsun$. In each panel, the plus sign indicates the values of the
central HOD model, the two dotted contours correspond to $\Delta\chi^2=1$
and 4, and the two solid contours are for $\Delta\chi^2=2.30$ and 6.17
(68.3\% and 95.4\% confidence levels for two parameters).
}
\end{figure*}

Figure~\ref{fig:mcmc_corr} shows
correlations between some pairs of HOD parameters, inferred from the MCMC run.
If galaxies are more centrally concentrated than dark matter inside halos, 
then the preferred width of $\PNM$ narrows to suppress one-halo pairs
in halos with small virial radii (Fig.~\ref{fig:mcmc_corr}$a$).
A wider $P(N|M)$ correlates, weakly, with a lower velocity bias
($\alpha_v < 1$ for $\omega > 1$; Fig.~\ref{fig:mcmc_corr}$b$),
since more small-scale pairs involve satellite galaxies, which make
larger contributions to the pairwise velocity dispersion.
There is a similar but stronger correlation with galaxy
concentration (Fig.~\ref{fig:mcmc_corr}$c$), since positive $\dlgc0$
allows more small-scale pairs to come from massive halos with
large velocity dispersions.  Finally, if the mean occupation at
high halo masses increases (higher $S_{15}$ in Fig.~\ref{fig:mcmc_corr}$d$),
then the mean halo mass at a given high multiplicity decreases, so
$\alpha_v$ greater than unity is favored to keep the {\it apparent}
virial mass $\Mviravg$ near its original value.

\subsection{Changing $\Omega_m$ with $\sigma_8$, $n_s$, and $\Gamma$ Fixed}
\label{sec:omegam}

Before considering general constraints in the $(\Omega_m,\sigma_8,n_s,\Gamma)$
parameter space in \S\ref{sec:general}, we examine constraints along 
several axes in this space: a pure change in $\Omega_m$, a pure change
in $\sigma_8$, ``cluster-normalized'' changes that preserve the value
of $\sigma_8\Omega_m^{0.5}$ at fixed $n_s$ and $\Gamma$, and linked
changes in $\Omega_m$, $\sigma_8$, and $\Gamma$ (at fixed $n_s$) that 
preserve the amplitude and slope of the halo mass function at high masses.
As discussed by \cite{Zheng02}, the halo population responds to these
changes in relatively simple ways, so we gain physical insight into 
the origin of the more general cosmological constraints by studying
the behavior along these axes.  These restricted constraints are also
relevant to the case where other parameters are determined by
independent data.  

\begin{figure}
\epsscale{1.2}
\plotone{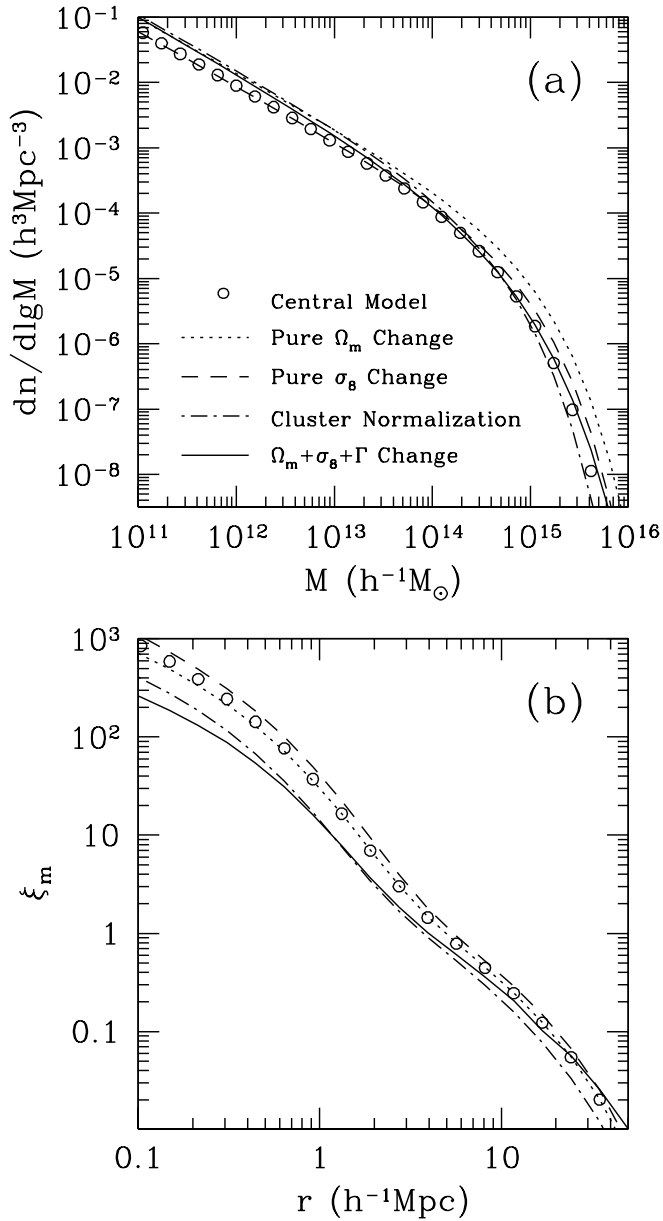}
\epsscale{1.0}
\caption[]{\label{fig:mf_xim}
Variations of ($a$) halo mass function and ($b$) mass correlation
function with cosmology. The fitting formula of
\citet{Sheth99} for the halo mass function and that of \citet{Smith03} for
the nonlinear mass power spectrum are used for the calculations.
In each panel, we show the results for the central cosmological model
($\Omega_m=0.3$, $\sigma_8=0.9$, and $\Gamma=0.2$; {\it open circles}),
a pure change in $\Omega_m$ to 0.5 ({\it dotted line}),
a pure change in $\sigma_8$ to 1.0 ({\it dashed line}), a cluster-normalized
model with $\Omega_m=0.5$ and $\sigma_8=0.7$ ({\it dot-dashed line}), and a
cluster-normalized model in which we also change the shape parameter to
$\Gamma=0.11$ to preserve the halo mass function as closely as possible
({\it solid line}).
}
\end{figure}

Figure~\ref{fig:mf_xim} illustrates the variation of the halo mass 
function $dn/d\log M$ (Fig.~\ref{fig:mf_xim}$a$) and the non-linear 
matter correlation function $\ximm(r)$ (Fig.~\ref{fig:mf_xim}$b$) along
these four axes.  Open circles show the central model.
Dotted lines show the effect of increasing $\Omega_m$ from 0.3 to 0.5,
with other parameters fixed.  Dashed lines show the effect of increasing
$\sigma_8$ from 0.9 to 1.0.  Dot-dashed lines show a cluster-normalized
change, with $\Omega_m$ increased to 0.5 and $\sigma_8$ decreased to 0.7
to keep $\sigma_8 \Omega_m^{0.5}$ fixed.  Solid lines show the
same $(\Omega_m,\sigma_8)$ combination, but with a corresponding change
in $\Gamma$ that keeps the slope of $dn/d\log M$ fixed at the cluster
mass scale $5\times 10^{14}\hMsun$ (see \citealt{Zheng02}).
In all cases we use the
\cite{Sheth99} mass function formula (our eq~[\ref{eqn:halomf}])
and the \cite{Smith03} prescription for computing the matter correlation
function.  Note that peculiar velocities of halos on large scales are
proportional to $\sigma_8\Omega_m^{0.6}$, similar to the scaling
of velocity dispersions of cluster mass halos \citep{Zheng02}.

The constraints for a pure $\Omega_m$ change are easiest to understand,
so we examine this case first and in the greatest detail.
As shown in Figure~\ref{fig:mf_xim}, a pure $\Omega_m$ change simply
shifts the mass function in proportion to $\Omega_m$.  The effect on
the matter correlation function is very small, arising entirely from
changes in the concentrations of halos.  Once halo masses are scaled in
proportion to $\Omega_m$, the spatial clustering of halos at a given
scaled mass is virtually unchanged, although the velocities change in
proportion to $\Omega_m^{0.6}$ \citep{Zheng02}.  

\begin{figure}
\epsscale{1.2}
\plotone{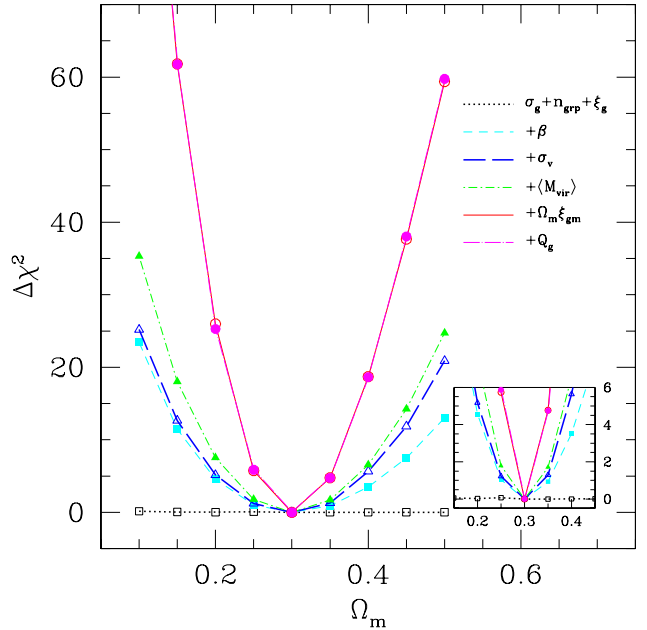}
\epsscale{1.0}
\caption[]{\label{fig:chi2_omega}
Values of $\Delta\chi^2$ as a function of $\Omega_m$, relative to the true
model with $\Omega_m=0.3$, in a cosmological model sequence where only
$\Omega_m$ is varied. As each set of observables is added in the analysis
(in the order indicated in the legend), the $\chi^2$ for each cosmological
model is minimized by solving the best-fit HOD parameters. Different line
(point) types show a sequence of $\Omega_m$ constraints from including more
and more complementary observables. The inset box shows the parts of the 
lines with $\Delta\chi^2<6$.
}
\end{figure}

\begin{figure*}
\epsscale{0.90}
\plotone{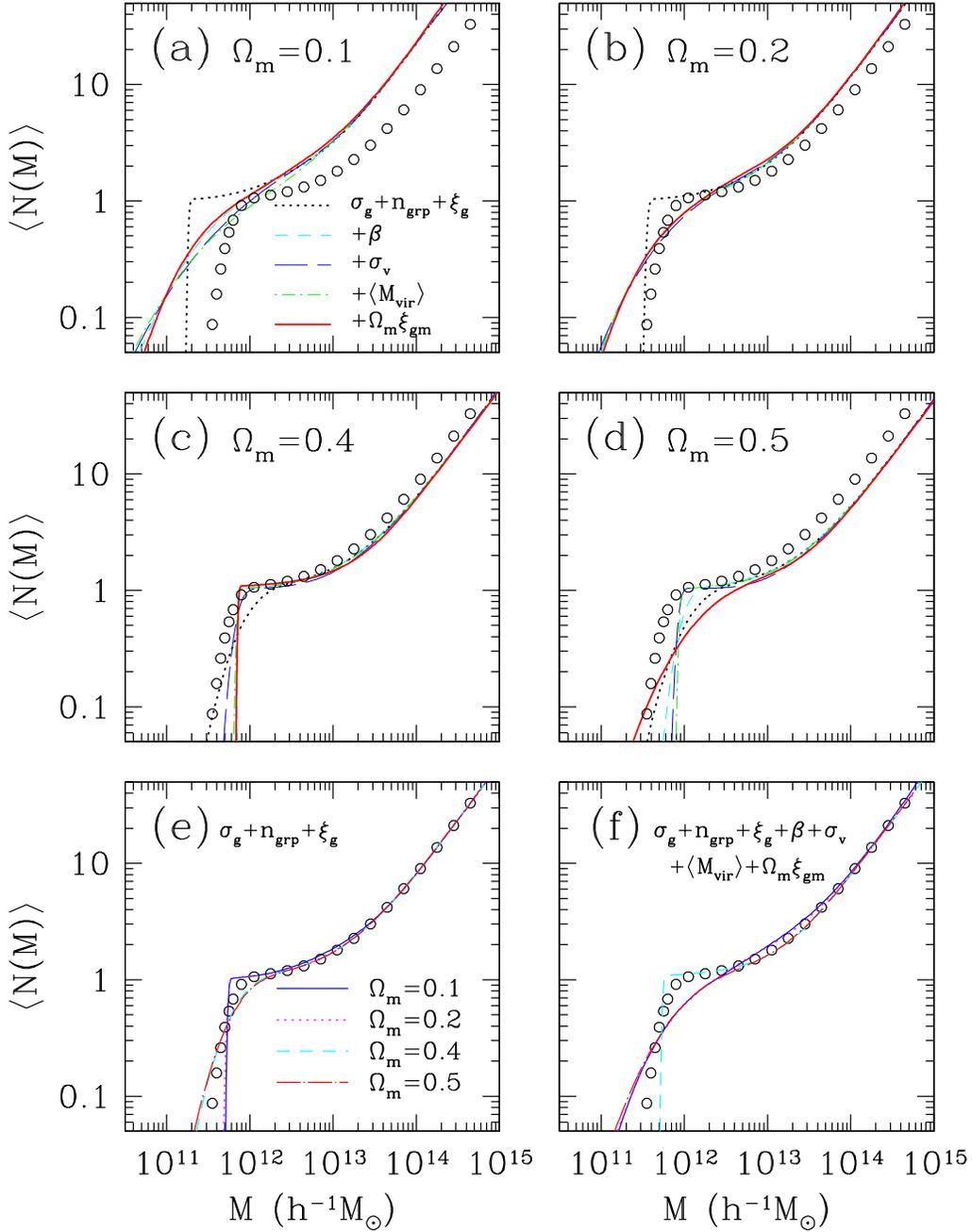}
\epsscale{1.00}
\caption[]{\label{fig:Navg1_omega}
($a$--$d$) Changes in $\Navg$ as new sets of observables are 
added, for cosmological models differing only in $\Omega_m$. The four panels
are for $\Omega_m=$0.1, 0.2, 0.4, and 0.5, respectively. For visual clarity,
we omit the line showing the effect of adding the reduced galaxy bispectrum
$Q_g$. ($e$) Best-fit $\Navg$ for different values of  
$\Omega_m$ if only spatial clustering observables are considered, after
scaling halo masses by $(\Omega_m/0.3)$.
($f$) Same scaled mass result
if other clustering observables are also included.
Open circles in each of the six panels show $\Navg$ for the
central cosmological model.
}
\end{figure*}

\begin{figure}
\epsscale{1.20}
\plotone{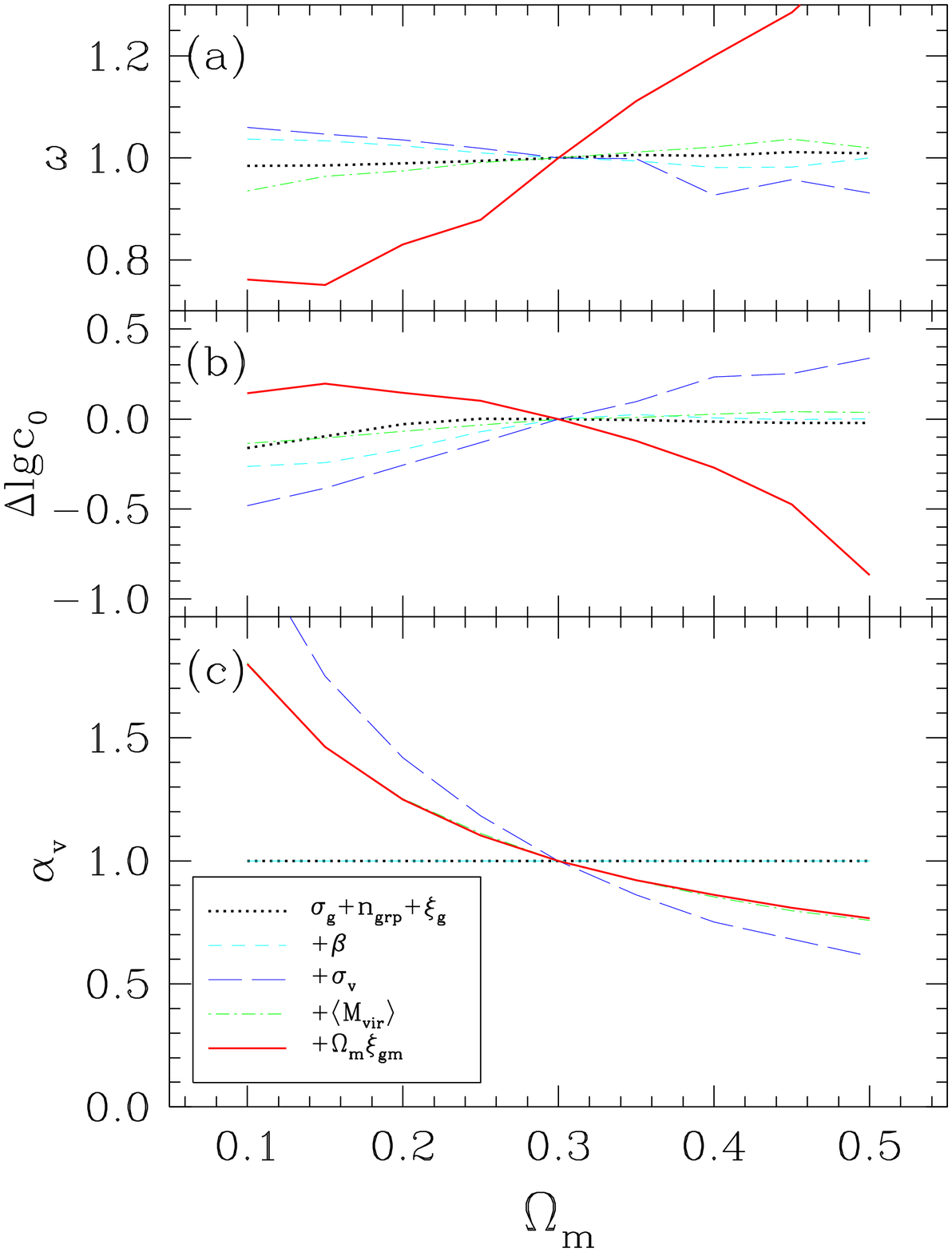}
\epsscale{1.00}
\caption[]{\label{fig:hodpar_omega}
Other parameters ($\omega$, $\Delta\log c_0$, and $\alpha_v$) of the best-fit
HODs as a function of $\Omega_m$ and the set of observables,
for cosmological models differing only in $\Omega_m$.
}
\end{figure}

Figure~\ref{fig:chi2_omega} shows $\Delta\chi^2$ for the best-fit HOD
model at each value of $\Omega_m$, as different observables are added
in succession.  Figures~\ref{fig:Navg1_omega} and~\ref{fig:hodpar_omega}
show the parameters of these best-fit HODs.  If we include only the
spatial clustering observables $\sigma_g(r)$, $\ngrp(\ge N)$, and
$\xigg(r)$, we obtain results virtually identical to those of the
central model by simply shifting the mass scale of the mean occupation
function in proportion to $\Omega_m$.  For these observables alone, 
we get $\Delta\chi^2 =0$ at all $\Omega_m$ ({\it horizontal dotted line
in} Fig.~\ref{fig:chi2_omega}).  Figure~\ref{fig:Navg1_omega}$e$ plots
the best-fit $\Navg$ as a function of the scaled halo mass 
$M\times (0.3/\Omega_m)$.  Because the constraints imposed by our
observables on the low end of $\Navg$ are weak, the cutoff profiles
of the scaled mass functions vary, but above $\meanN \sim 1.2$
they are nearly indistinguishable.  Thus, the spatial clustering
observables effectively fix $\Navg$ in terms of scaled halo masses,
and only small further adjustments can be accommodated when other
observables are introduced. As shown in Figure~\ref{fig:hodpar_omega}, 
the best-fit models have $\omega=1$ just like the central model HOD,
and they have $\dlgc0 \approx 0$, although there is a slight reduction
in halo concentration at low $\Omega_m$ to compensate for the higher
concentrations of the halos themselves.

The $\Omega_m$ degeneracy breaks as soon as we include dynamical measures
that are sensitive to the absolute mass or velocity scale of halos.
The first such measure that we add is $\beta \equiv \Omega_m^{0.6}/b_g$.
For low $\Omega_m$, the cutoff of $\Navg$ smooths out so that more massive
galaxies come from lower mass halos with lower bias factors 
({\it dashed lines in} Figs.~\ref{fig:Navg1_omega}$a$ 
and \ref{fig:Navg1_omega}$b$).
The $\PNM$ distribution broadens slightly ($\omega >1$ in 
Fig.~\ref{fig:hodpar_omega}$a$) to compensate for a drop in one-halo
pairs, and concentrations drop ($\dlgc0 < 0$ in Fig.~\ref{fig:hodpar_omega}$b$)
so that these pairs move to larger separations.  However, within the
constraints imposed by the multiplicity function, the correlation 
function, and the galaxy number density, there is virtually no room
to change the galaxy bias factor in a way that compensates changes
in $\Omega_m$.  For $\Omega_m$ in the range $0.2-0.4$, the $\Delta\chi^2$
values in Figure~\ref{fig:chi2_omega} are almost equal to those
expected for constant $b_g$, given our adopted 10\% uncertainty in $\beta$
(and hence $\Omega_m^{0.6}$).

When we add galaxy pairwise velocity dispersions as observables, the
velocity bias parameter $\alpha_v$ adjusts to compensate for changes
in $\Omega_m$ (Fig.~\ref{fig:hodpar_omega}$c$, {\it long-dashed line}).
However, while velocity bias can reduce velocity dispersions within
halos, it cannot affect the pairwise velocities of the halos themselves,
which scale as $\Omega_m^{0.6}$.  For high $\Omega_m$, $\alpha_v$
drops by more than the factor $(0.3/\Omega_m)^{1/2}$ that would keep
{\it internal} dispersions fixed, in an effort to compensate for
the higher halo velocities.  This change would drive the pairwise
dispersion too low on small scales, where the one-halo term dominates,
so halo concentrations increase, allowing more of the small separation
pairs to come from massive halos.  In response to these higher concentrations,
the width parameter $\omega$ drops, reducing the one-halo contribution
to $\xigg(r)$.  Despite the freedom introduced by velocity bias, the
combination of pairwise dispersions at small and large scales adds
significant discriminatory power on the value of $\Omega_m$.  For
$\Omega_m=0.4$, for example, $\Delta\chi^2$ increases from $\sim 3.5$
($\beta$ alone) to $\sim 5.6$, with the two dynamical measurements
producing most of the discrepancy.

When we add the $\Mviravg$ observables, the velocity bias is forced close
to the value $\alpha_v=(0.3/\Omega_m)^{1/2}$ that keeps these apparent
virial masses fixed (Fig.~\ref{fig:hodpar_omega}$c$, {\it dot-dashed line,
nearly obscured by solid line}).  Since there is no longer room
to improve the large-scale pairwise dispersion by ``overcompensating''
with $\alpha_v$, the concentration and width parameters return to
the values preferred by the spatial clustering observables
(note the similarity of dotted and dot-dashed lines in 
Figs.~\ref{fig:hodpar_omega}$a$ and \ref{fig:hodpar_omega}$b$, 
and the difference between these
lines and the long-dashed lines).  Although $\alpha_v$ can 
exactly compensate a pure $\Omega_m$ shift for the $\Mviravg$
observables, the match to the pairwise dispersion becomes worse
as a result, and $\Delta\chi^2$ increases slightly.

The galaxy-galaxy lensing observable $\Omega_m\xigm(r)$ provides a
diagnostic of host halo masses that is unaffected by velocity bias.
If $\Navg$ stays fixed as a function of scaled mass, then a pure
$\Omega_m$ change simply scales $\Omega_m\xigm(r)$ in proportion
to $\Omega_m$ (eq.~\ref{eqn:xigm1h}).  The effect of higher $\Omega_m$
can be partly compensated by reducing halo concentrations
(Fig.~\ref{fig:hodpar_omega}$b$, {\it solid line}), 
which dilutes the contribution from satellite galaxy and matter 
particle pairs.
This decrease drives
an increase in $\omega$ (Fig.~\ref{fig:hodpar_omega}$a$) to restore
lost one-halo galaxy pairs at small separations 
(note that $\xigm$ itself is independent of $\omega$).
However, altering concentrations has limited power to compensate
for a shift in the halo mass function, and $\Omega_m\xigm(r)$ 
measurements add substantially to $\Delta\chi^2$.
Measurements in the linear or near-linear regime, not considered here,
would add still more discriminatory power.

Finally, large-scale bispectrum measurements have little sensitivity
to a pure $\Omega_m$ shift (Fig.~\ref{fig:chi2_omega}, {\it long-dashed line}).
This statistic is mainly a diagnostic for the galaxy bias factor $b_g$.
Along this model sequence, the amplitude of dark matter clustering
does not change, and the galaxy bias factor is therefore determined
by the observed spatial clustering, independent of $\Omega_m$.
The bispectrum will become important when we consider sequences with
changing $\sigma_8$.

Overall, these galaxy clustering observables have great power to
constrain any pure shift in $\Omega_m$.  A model with $\Omega_m \neq 0.3$
has a likelihood $\exp(-\Delta\chi^2/2)$ relative to the central model,
so even a model with $\Delta\chi^2 = 4.6$ is disfavored by a factor
of 10.  Including all observables, we find $\Delta\chi^2 \sim 5.8$
and 5.0 for $\Omega_m=0.25$ and 0.35, respectively, implying that 
$\Delta\Omega_m = 0.05$ can be firmly ruled out.  These constraints
weaken significantly if we omit the weak-lensing observables with their
direct sensitivity to mass scales.  Still, even with $\beta$ and
$\sigma_v(r)$ as the only dynamical observables, the $\Omega_m=0.2$
and 0.4 models are rejected with $\Delta\chi^2 > 5$.

\subsection{Changing $\sigma_8$ with $\Omega_m$, $n_s$, and $\Gamma$ Fixed}
\label{sec:puresig8}

\begin{figure}
\epsscale{1.2}
\plotone{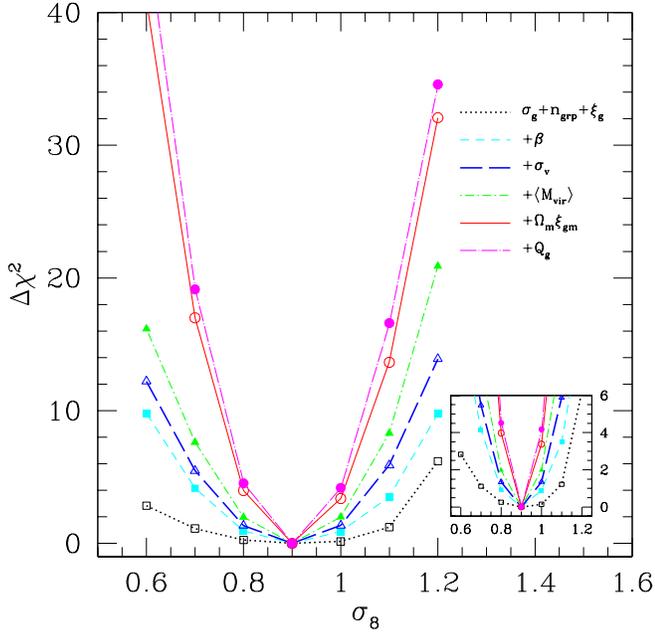}
\epsscale{1.0}
\caption[]{\label{fig:chi2_sig8}
Values of $\Delta\chi^2$ as a function of $\sigma_8$, relative to the true
model with $\sigma_8=0.9$, in a cosmological model sequence where only
$\sigma_8$ is varied. The format is similar to Fig.~\ref{fig:chi2_omega}.
}
\end{figure}

\begin{figure*}
\plotone{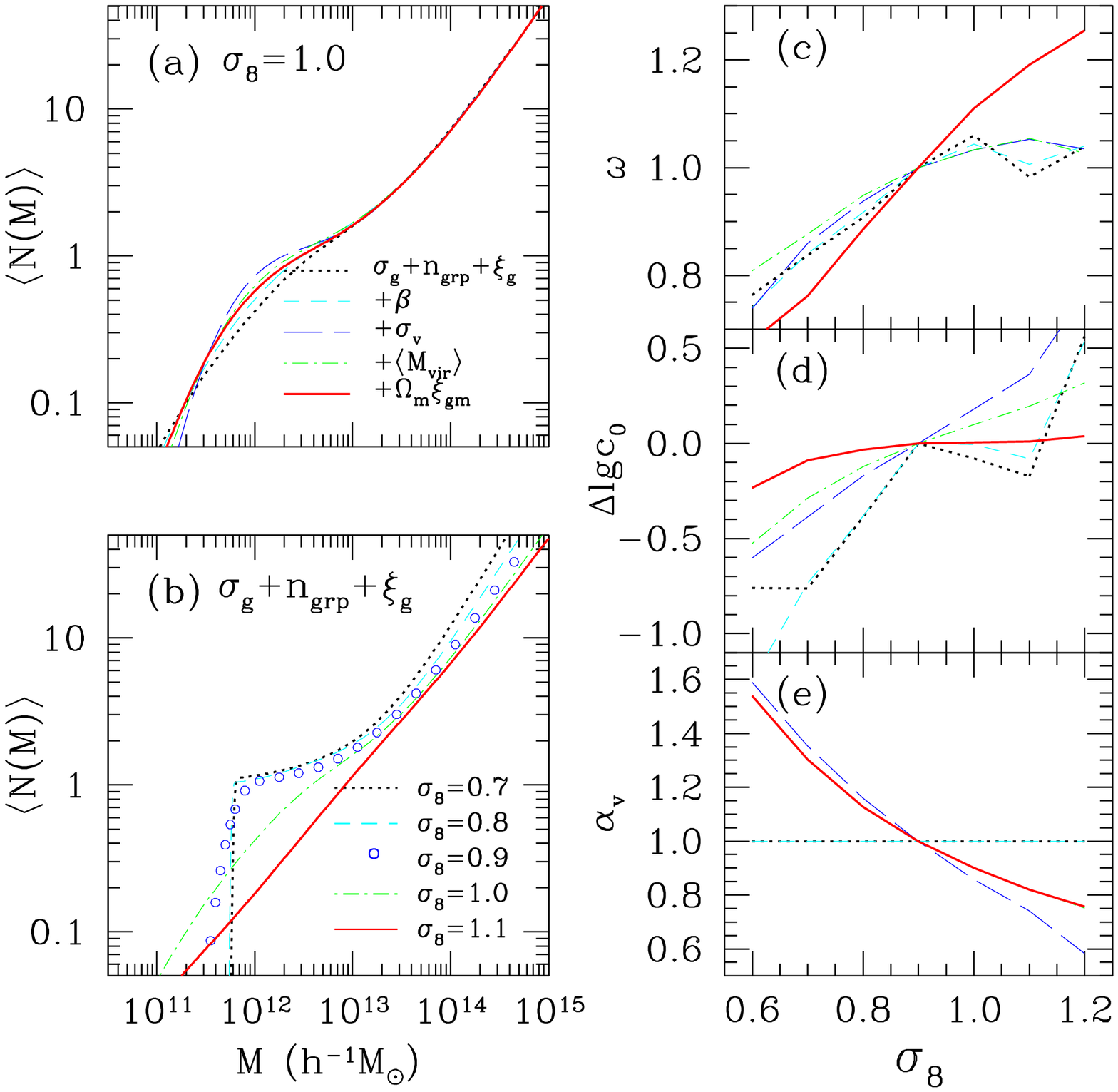}
\caption[]{\label{fig:hodpar_sig8}
Best-fit HODs for cosmological models differing only in $\sigma_8$.
($a$) $\Navg$ curves for the $\sigma_8=1.0$ cosmological model as
different observables are included. (b) Best-fit $\Navg$
curves for a range of $\sigma_8$ values when only spatial clustering
observables are considered. The central model ({\it open circles}) has
$\sigma_8=0.9$. ($c$, $d$, $e$) Other parameters ($\omega$,
$\Delta\log c_0$, and $\alpha_v$) of the best-fit HODs as a function of
$\sigma_8$ for the same set of observables shown in ($a$).
In ($e$), the dot-dashed line is obscured by the solid line.
}
\end{figure*}

We now consider models that differ only in $\sigma_8$, the amplitude
of the linear matter power spectrum, with $\Omega_m$ and the shape
of $P(k)$ held fixed.  The dashed lines in Figure~\ref{fig:mf_xim}
show the effect of raising $\sigma_8$ from 0.9 to 1.0.  The amplitude
of the matter correlation function increases at all scales, and the 
space density of high-mass halos ($M \ga 10^{14}\hMsun$) rises,
while the space density of galaxy
mass halos ($M \sim 10^{12}-10^{13}\hMsun$) is nearly unchanged.
Figure~\ref{fig:chi2_sig8} shows $\Delta\chi^2$ for the best-fit
HOD models along the changing $\sigma_8$ sequence, and 
Figure~\ref{fig:hodpar_sig8} shows the properties of these
best-fit models.

Figure~\ref{fig:hodpar_sig8}$b$ shows the response of $\Navg$ to
$\sigma_8$ when the spatial clustering observables $\sigma_g$,
$\xigg$ and $\ngrp$ are the only constraints.  Higher $\sigma_8$
leads to lower $\Navg$ at high masses, so that $\ngrp$ remains
fixed despite an increased halo abundance.  This change reduces
the galaxy bias factor $b_g$, but on its own it is not enough
to compensate for the increased matter clustering amplitude.
The $\Navg$ cutoff therefore spreads to lower masses, putting
more isolated galaxies in low bias halos while retaining the
galaxy number density $n_g$.  The $\PNM$ width parameter rises
to $\omega>1$ to restore small-scale pairs lost when the number
of halos in the $\Navg \sim 1-2$ range is reduced
(Fig.~\ref{fig:hodpar_sig8}$c$, {\it dotted line}).

For spatial clustering observables alone, these HOD changes
effectively compensate for changes in the halo population,
and the $\Delta\chi^2$ curve is nearly flat in the vicinity
of the central model (Fig.~\ref{fig:chi2_sig8}, {\it dotted line}).
However, these observables pin down $\Navg$ fairly precisely,
leaving little room for adjustment when dynamically sensitive
observables are added as constraints.  
The $\Delta\chi^2$ from $\beta = \Omega_m^{0.6}/b_g$ is
close to that expected for our 10\% error bar and the scaling
$b_g \propto 1/\sigma_8$ implied by fixed galaxy spatial clustering.
The velocity bias parameter $\alpha_v$ can partly compensate for
changes in halo velocities and internal dispersions, but it cannot
fix the small- and large-scale values of $\sigma_v(r)$ simultaneously
(see discussion in \S\ref{sec:omegam}),
so $\dchi2$ rises when $\sigma_v$ is added.
Adding $\Mviravg$ effectively pins down $\alpha_v$ to the
value needed to keep the apparent virial masses fixed, making
the overall fit of $\sigma_v(r)$ worse and further increasing $\dchi2$.
The $\Omega_m\xigm(r)$ observable is unaffected by velocity bias,
and the small adjustments still allowed in $\Navg$, $\dlgc0$,
and $\omega$ do little to counter the changes in halo masses.
Finally, the galaxy bispectrum increases $\dchi2$ through its 
direct sensitivity to $b_g$.  With all observables considered
together, $\dchi2$ exceeds 4.0 for $\Delta\sigma_8 = \pm 0.1$,
and it rises rapidly for larger changes.

\subsection{Cluster-normalized Models}
\label{sec:clnorm}

The observed abundance of massive galaxy clusters effectively
constrains the amplitude of the halo mass function at a scale
$M \sim 5\times 10^{14}\hMsun$.  The halo abundance at this
scale is an increasing function of $\Omega_m$ and $\sigma_8$,
and the cluster normalization constraint is usually expressed
in the form $\sigma_8\Omega_m^q={\rm const.}$, with
$q \sim 0.5$ 
(e.g., \citealt{White93,Eke96,Pierpaoli01}). Here,
as in \citet{Zheng02}, we consider a sequence of cluster-normalized 
cosmological models that satisfy $\sigma_8=0.9(\Omega_m/0.3)^{-0.5}$, with
the shape of $P(k)$ held fixed.
The dot-dashed lines in Figure~\ref{fig:mf_xim} show the effect
of simultaneously increasing $\Omega_m$ to 0.5 and decreasing
$\sigma_8$ to 0.70.  As expected, the halo mass function stays
the same at $M \sim 5\times 10^{14}\hMsun$, but relative to the
fiducial model there are more low-mass halos and fewer halos
with $M \ga 10^{15}\hMsun$.  The matter correlation function is
lower at all scales because of the lower value of $\sigma_8$.

\begin{figure}
\epsscale{1.2}
\plotone{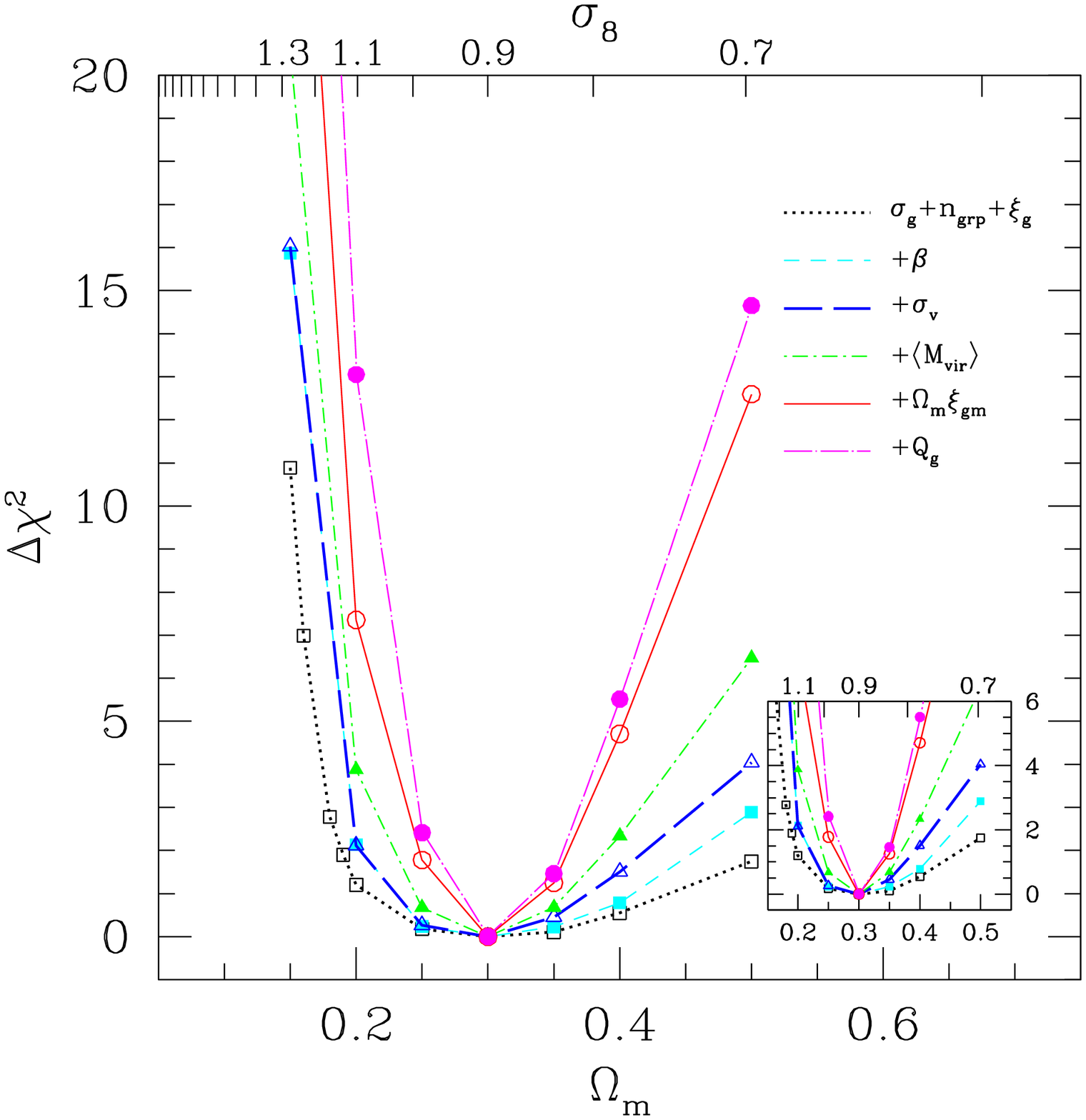}
\epsscale{1.0}
\caption[]{\label{fig:chi2_clnorm}
Same as Fig.~\ref{fig:chi2_omega}, but for a cluster-normalized model sequence
where $\sigma_8\Omega_m^{0.5}$ is held fixed as $\Omega_m$ is varied.
The corresponding values of $\sigma_8$ are marked on the top axis.
}
\end{figure}

\begin{figure*}
\plotone{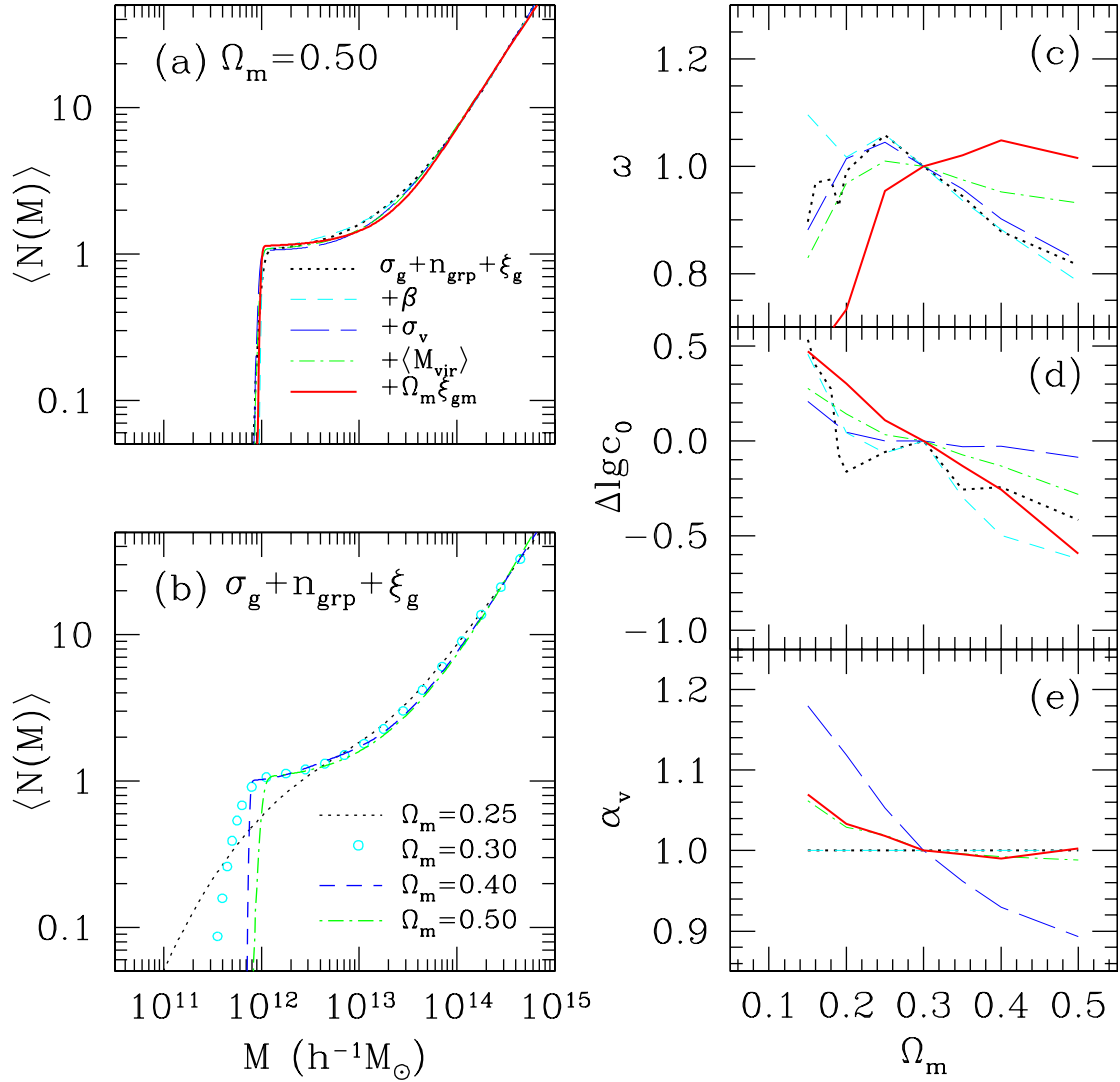}
\caption[]{\label{fig:hodpar_clnorm}
Similar to Fig.~\ref{fig:hodpar_sig8}, but for the cluster-normalized
model sequence.
}
\end{figure*}

\begin{figure}
\epsscale{1.2}
\plotone{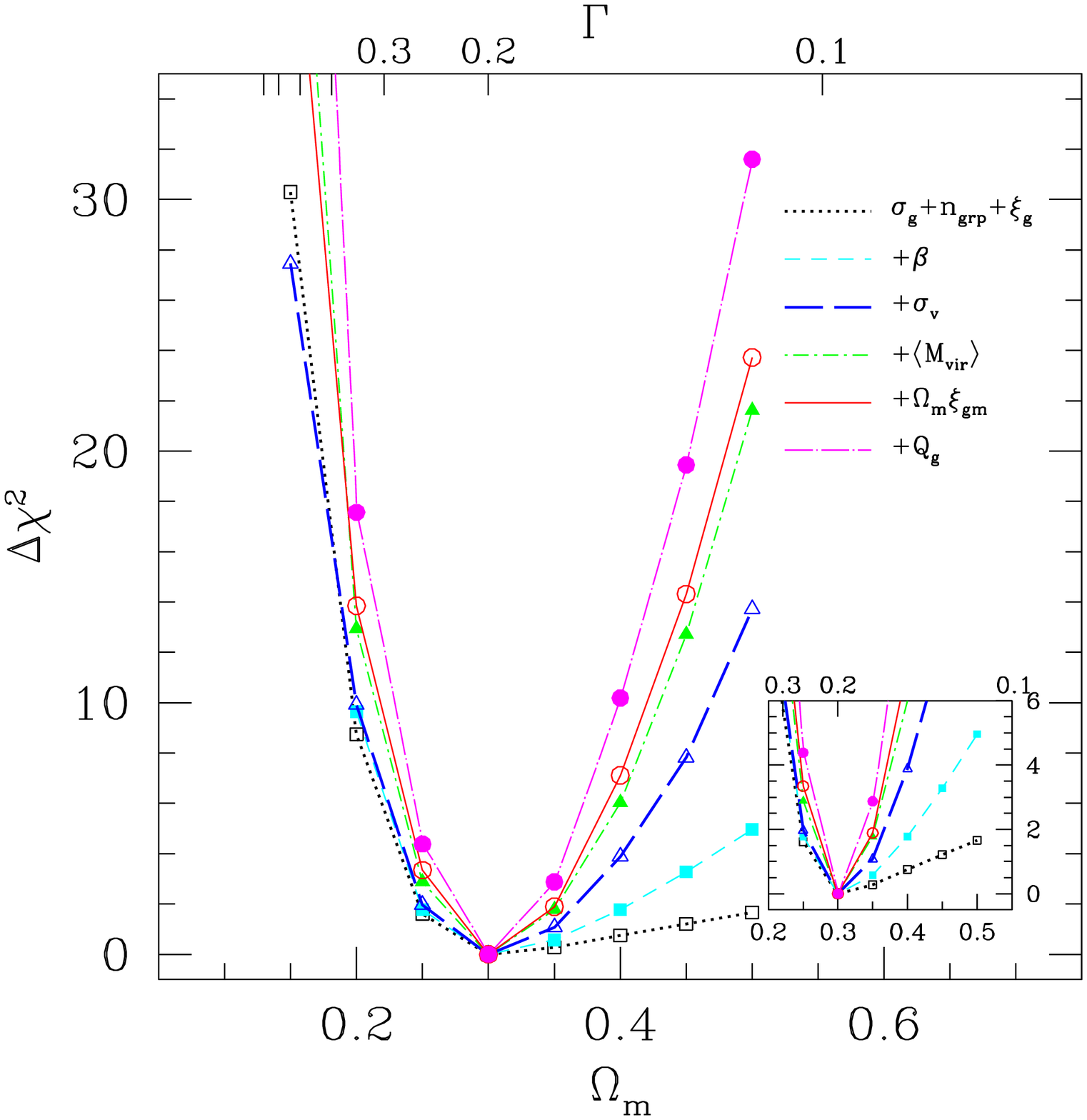}
\epsscale{1.0}
\caption[]{
\label{fig:chi2_mfmatch}
Same as Fig.~\ref{fig:chi2_omega}, but for a model sequence in which
$\sigma_8$ and $\Gamma$ are varied as a function of $\Omega_m$ to preserve
the halo mass function as closely as possible. The corresponding values of
the shape parameter $\Gamma$ are marked at the top axis.
}
\end{figure}

Figures~\ref{fig:chi2_clnorm} and~\ref{fig:hodpar_clnorm} show the
$\dchi2$ and parameters of the best-fit HOD models along the 
cluster-normalized sequence, in the same format as 
Figures~\ref{fig:chi2_sig8} and~\ref{fig:hodpar_sig8}.
The multiplicity function fixes the high-mass end of $\Navg$
in a way that is nearly independent of $\Omega_m$, although there
are slight variations in slope that reflect the changing slopes
of the halo mass function (Fig.~\ref{fig:hodpar_clnorm}$b$).
The higher abundance of low-mass halos for high $\Omega_m$
requires an increase in the cutoff mass to keep the 
galaxy number density constant.  The tightened mass range 
for single-galaxy halos leads to a larger one-halo contribution
to $\xigg(r)$, which is compensated by decreasing the width
parameter $\omega$ and the concentration parameter $\dlgc0$.
Although the matter correlation amplitude is reduced for high
$\Omega_m$ and low $\sigma_8$, the bias of halos at fixed
mass is higher, and these two effects conspire to keep the
large-scale amplitude of the galaxy correlation function nearly
constant over a substantial range of $\Omega_m$.
For spatial clustering observables alone, $\dchi2$ is nearly
flat over the range $0.2 \leq \Omega_m \leq 0.5$.

Because cluster normalization keeps the mass and velocity scale
of halos roughly fixed, the addition of dynamically sensitive 
observables makes relatively little difference to best-fit
HOD parameters, and $\dchi2$ increases more slowly than it does
for pure $\Omega_m$ or pure $\sigma_8$ changes (note the smaller
vertical scale of Figure~\ref{fig:chi2_clnorm} relative to 
Figures~\ref{fig:chi2_omega} and~\ref{fig:chi2_sig8}).  
The detailed changes in $\omega$, $\dlgc0$, and $\alpha_v$
as new observables are added mostly reflect the changes in
the cutoff of $\Navg$ and corresponding changes in the
fraction of single-occupancy halos.  The $\dchi2$ curves
are asymmetric in $\Omega_m$ because of the non-linear
relation between $\Omega_m$ and $\sigma_8$.  
For example, cluster normalization requires a 13\% reduction
in $\sigma_8$ to 0.78 at $\Omega_m=0.4$ but a 22\% increase
to $\sigma_8=1.10$ at $\Omega_m=0.2$, and the still larger
$\sigma_8$ values required at lower $\Omega_m$ are easily 
ruled out.  The bispectrum constraint is especially useful
in distinguishing cluster-normalized models because it
responds to $b_g$ independent of $\Omega_m$.  Despite the
compensating effects of $\sigma_8$ and $\Omega_m$ in cluster-normalized
models, the full set of observables yields $\dchi2 \sim 5.5$ for
$\Omega_m=0.4$ and $\dchi2 \sim 13.1$ for $\Omega_m = 0.2$.

As shown by \cite{Zheng02}, one can construct cluster-normalized
model sequences that match the amplitude {\it and slope} of the 
halo mass function at cluster scales by changing the power spectrum
shape parameter in concert with $\Omega_m$ and $\sigma_8$.
The solid lines in Figure~\ref{fig:mf_xim} show
a model with $\Omega_m=0.5$, $\sigma_8=0.71$, and
$\Gamma=0.11$.  The halo mass function is nearly identical
to the central model's for $M \ga 10^{13.5}\hMsun$, although
it is higher at low halo masses.  The
{\it shape} of the matter correlation function is significantly different
because of the different shape of $P(k)$.

Figure~\ref{fig:chi2_mfmatch} shows $\dchi2$ as a function
of $\Omega_m$ along this model sequence.  The required values
of $\sigma_8$ and $\Gamma$ are shown in 
Figure~13 of \cite{Zheng02}, and the values of $\Gamma$ are
also marked on the top axis of our Figure~\ref{fig:chi2_mfmatch}.
As speculated by \cite{Zheng02}, matching the slope of the halo
mass function by changing the shape of $P(k)$ does more harm than
good: $\dchi2$ values at a given $\Omega_m$ are always higher than
they are for the fixed $P(k)$, cluster-normalized models shown
in Figure~\ref{fig:chi2_clnorm}.  With all observables included,
models with $\Omega_m=0.2$ and $\Omega_m=0.4$ have
$\dchi2 \sim 17.6$ and $\dchi2 \sim 10.2$, respectively.

\subsection{More General Cases}
\label{sec:general}

The four model sequences considered in 
Figures~\ref{fig:chi2_omega}, \ref{fig:chi2_sig8},
\ref{fig:chi2_clnorm}, and \ref{fig:chi2_mfmatch}
trace four different curves through the 
$(\Omega_m, \sigma_8, \Gamma)$ space of cosmological
parameters.  We now examine the constraints and 
degeneracies in this parameter space in more general
terms.  The computational expense of finding the best-fit,
10-parameter HOD model for each cosmological model makes
a comprehensive study of the full 3D space
difficult.  However, the fact that changing the power spectrum
shape to match the halo mass function makes models easier 
to distinguish suggests that the ``$\Gamma$'' dimension is
largely decoupled from the $\Omega_m-\sigma_8$ plane.
Furthermore, the galaxy power spectrum shape can be measured
directly on large scales (e.g., \citealt{Tegmark04a,Cole05,Tegmark06}),
and we have not incorporated these direct constraints into our 
observables.

\begin{figure*}
\plotone{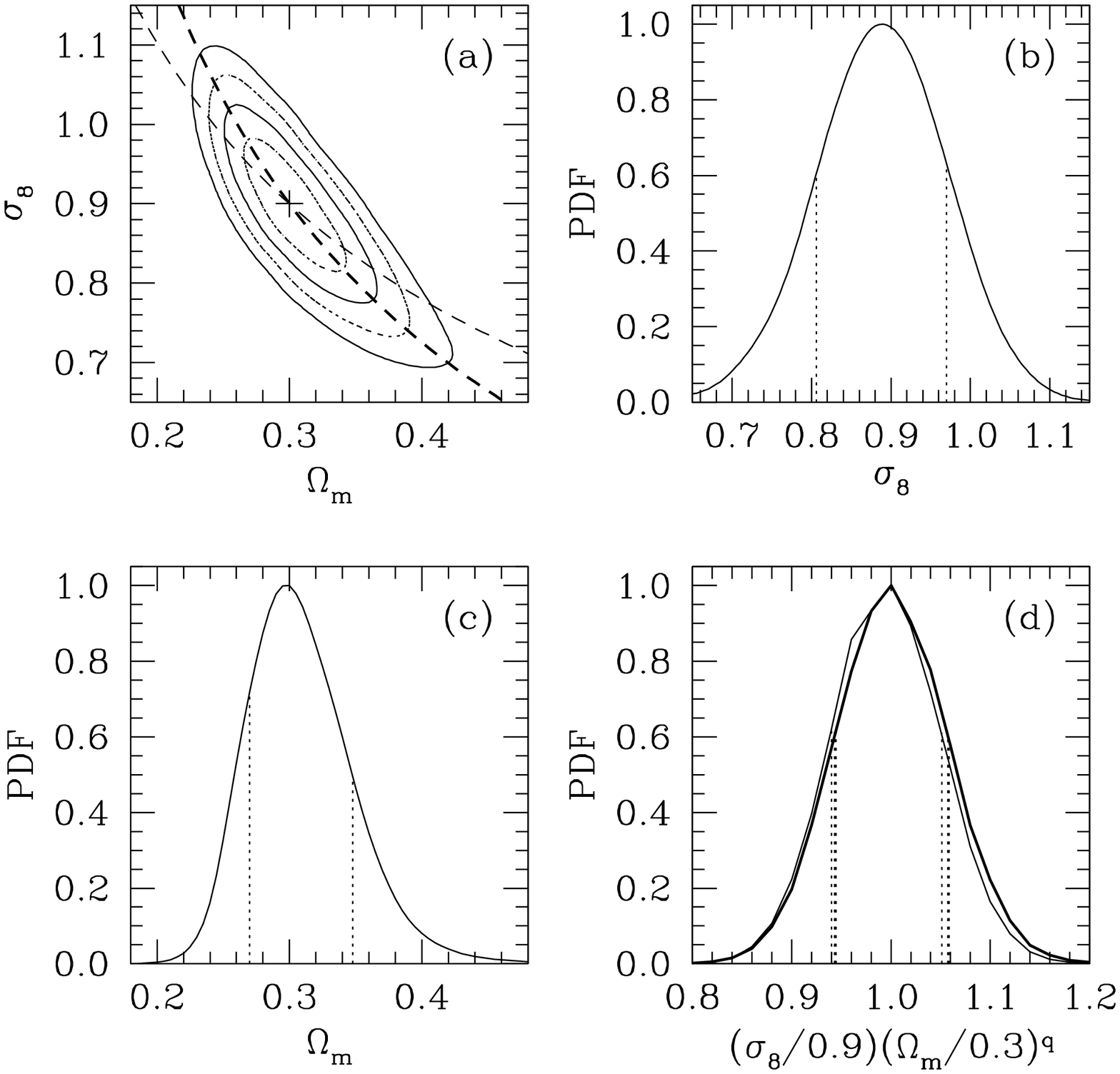}
\caption[]{\label{fig:contour}
Constraints on $\Omega_m$ and $\sigma_8$ (with $\Gamma$ fixed)
from galaxy clustering data. ($a$) Contour plot.
The two dotted contours correspond to $\Delta \chi^2=1$ and 4, and the
two solid ones are for $\Delta \chi^2=2.30$ and 6.17 (i.e., about the 
68.3\% and 95.4\% confidence levels for two parameters). The two dashed 
lines represent $(\sigma_8/0.9)(\Omega_m/0.3)^q=1$ with $q=0.5$ ({\it thin})
and $q=0.75$ ({\it thick}). The other three panels show marginalized 
likelihoods on $\sigma_8$, $\Omega_m$, and $\sigma_8\Omega_m^q$, respectively.
Vertical dotted lines in each of these three panels mark the central
68.3\% of the distribution (i.e., the 1-$\sigma$ range). In ($d$), thin 
and thick lines are for $q=0.5$ and 0.75, respectively.
}
\end{figure*}

To begin, therefore, we fix $\Gamma=0.2$ and map out $\Delta\chi^2$
contours in the $\Omega_m-\sigma_8$ plane, finding
the best-fit HOD parameters 
that minimize the values of $\Delta\chi^2$ at each point on the 
cosmological parameter grid.  We include all the
observables discussed in \S\ref{sec:observables} and 
incorporated in our earlier constraints.
The solid contours in Figure~\ref{fig:contour}$a$ show $\dchi2=2.30$ 
and $\dchi2=6.17$, corresponding to 68.3\% and 95.4\%
confidence intervals for two parameters.
The most degenerate direction seems to be
$\sigma_8\Omega_m^q={\rm const.}$ with $q\sim0.7-0.8$, which deviates
from the degeneracy axis of cluster normalization
($q\sim0.5$) or large-scale redshift-space distortions ($q\sim 0.6$).
This deviation reflects the steeper scaling favored by
the galaxy-galaxy lensing constraint, which has $q\sim 1$ even
on these non-linear scales (\citealt{Yoo06}), and by pairwise
velocity dispersions with $\alpha_v$ as an adjustable parameter.

Dotted contours in this panel show $\dchi2=1$ and 4, from which one can
read projected $1\sigma$ and $2\sigma$ constraints on the 
individual parameters $\sigma_8$ and $\Omega_m$.
The remaining panels show the 1D likelihoods for
these parameters and the combination $\sigma_8\Omega_m^q$,
inferred by integrating the likelihood 
function $\exp(-\Delta\chi^2/2)$ over each bin of the relevant quantity.
Under our assumption of 10\% fractional error in each observable, 
the 1-$\sigma$ uncertainty in the matter density parameter $\Omega_m$ 
is about 11\%. The amplitude $\sigma_8$ of the matter power spectrum 
has a 1-$\sigma$ constraint that is better than 10\%. The 1-$\sigma$ 
uncertainty in the best constrained quantity, $\sigma_8\Omega_m^q$ 
with $q =0.75$, is at the 5\% level.  If we force $q$ to 0.5, the
precision is almost equally good.

What constraints could we infer on $\Gamma$ from the galaxy clustering 
measurements considered in our observable set?
To address this question in approximate terms, we fix 
$\Omega_m$ and $\sigma_8$ to the central model values of 0.3 and 0.9
and vary $\Gamma$ alone, again finding best-fit HOD parameters
and minimum $\dchi2$.  We obtain $\dchi2=1$ at $\Gamma=0.183$, 0.219
and $\dchi2=4$ at $\Gamma=0.167$, 0.240.  The 10\% ($1\sigma$) constraint on 
$\Gamma$ is similar to that found along our matched halo mass
function sequence ($\dchi2=1$ at $\Gamma=0.177$, 0.223
and $\dchi2=4$ at $\Gamma=0.161$, 0.250; Figure~\ref{fig:chi2_mfmatch}), 
supporting our conjecture that the ($\Omega_m,\sigma_8$) constraints are
mostly orthogonal to the $P(k)$ shape constraints.
With our 30 observables and assumption of 10\%, independent
fractional errors, therefore, the full $\chi^2$ surface
approximately follows the contours of Figure~\ref{fig:contour}
with an orthogonal Gaussian of roughly 10\% $1\sigma$ width 
in the $\Gamma$ direction.

As discussed in the introduction, traditional approaches to 
testing cosmological models with large-scale structure focus
on the perturbative regime, where it is reasonable to adopt
linear theory and linear bias for describing the galaxy power
spectrum and redshift-space distortions and second-order perturbation
theory and a quadratic bias model for describing the galaxy bispectrum.
It is interesting to compare our HOD-based constraints to those 
that could be derived from the same observables with the 
perturbative approach.  Specifically, we use the $\sigma_g$,
$\beta$, and $Q_g$ observables adopted previously, with the same 10\%
uncertainties.  We employ a two-parameter quadratic bias model, and
we calculate the observables for a given cosmology and bias
using $\sigma_g = b\sigma$, $\beta = \Omega_m^{0.6}/b$, and
$Q_g = Q_m/b+b_2/b^2$, where $\sigma$ and $Q_m$ are the rms
fluctuations and reduced bispectrum for the dark matter.
We examine an ($\Omega_m,\sigma_8$) grid with $\Gamma=0.2$ and find
the values of $b$ and $b_2$ that minimize $\dchi2$ for each
cosmological model.

\begin{figure*}
\plotone{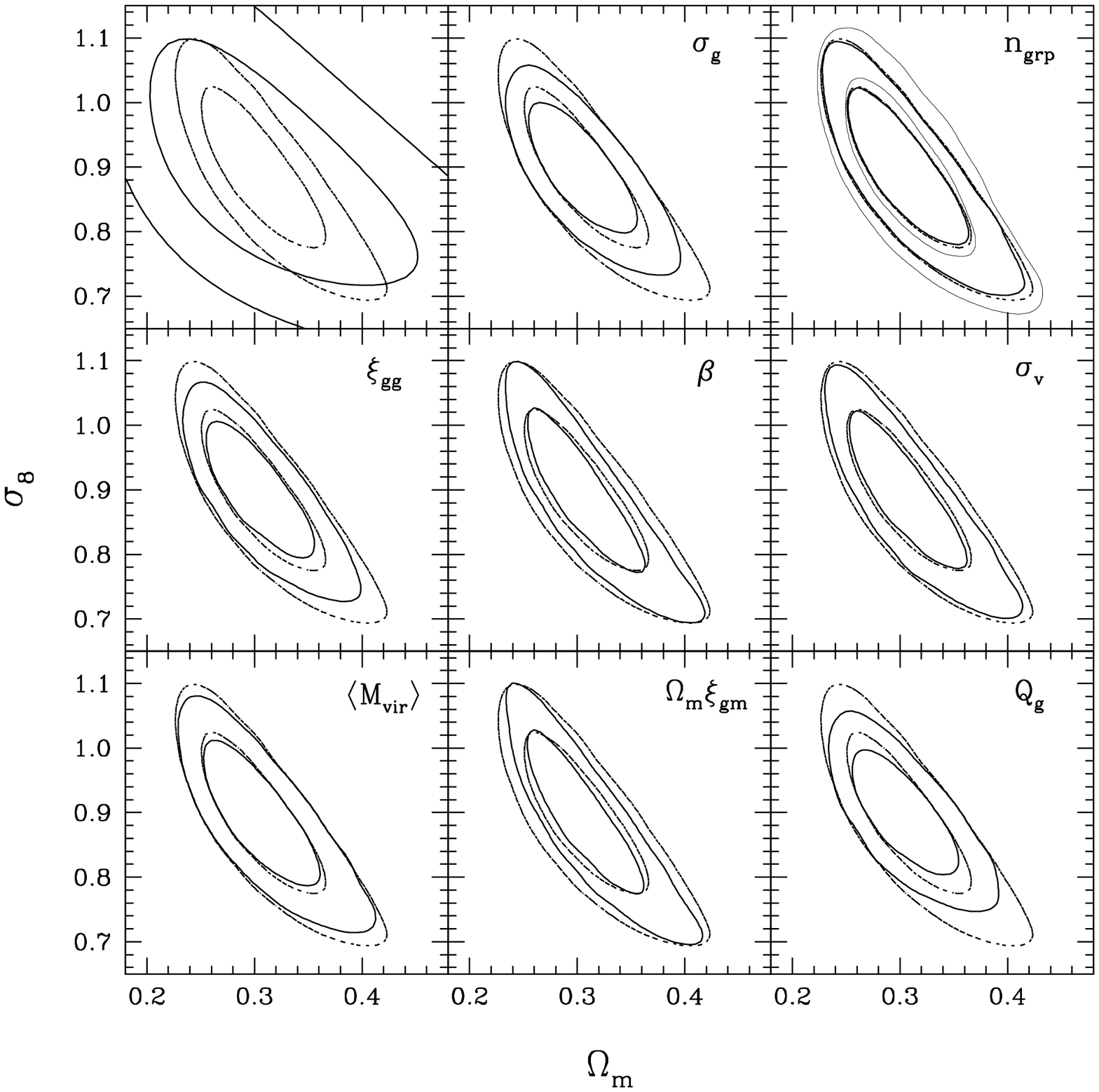}
\caption[]{\label{fig:chgerr}
Constraints on $\Omega_m$ and $\sigma_8$ (with $\Gamma$ fixed)
from galaxy clustering data. Dotted contours in each panel are the same
as the solid contours in Fig.~\ref{fig:contour}, i.e., 68.3\% and 95.4\%
confidence levels by assuming 10\% fractional error in each of the 30
observables. Solid contours in the top-left panel are corresponding
constraints using $\sigma_g$, $\beta$, and $Q_g$ and applying the
perturbative regime model (see text). Solid contours in each of the
other panels are constraints from HOD modeling by changing the fractional
errors to be 5\% for the set of observables indicated in the panel.
In the panel for $\ngrp$, we also show the case with the fractional
errors in the multiplicity function changed to 50\% ({\it thin solid 
contours}).
}
\end{figure*}

The top left panel of Figure~\ref{fig:chgerr} compares our
HOD-based constraints (repeated from Figure~\ref{fig:contour})
to those obtained from the perturbative analysis.
The principal degeneracy axis for the latter constraints is
$\sigma_8\Omega_m^{0.6}={\rm const.}$, since $\beta$ constrains
$\Omega_m^{0.6}/b$
and $\sigma_g$ constrains $b\sigma_8$.
The bispectrum constraint on $b$ breaks this degeneracy and produces
closed contours.  However, the constraints from the full HOD-based
analysis of all observables are much tighter than the constraints
from the smaller set of observables that can be modeled by the
perturbative approach alone.  Of course, the perturbative cosmology
constraints can be improved by including more observables,
such as large-scale weak-lensing measurements or CMB anisotropies,
but these observables will also improve the HOD-based constraints.
Thus, despite the need for a more complex, multi-parameter model of
galaxy bias, the ability of the HOD approach to model intermediate
and small-scale clustering leads to substantial improvements
in cosmological power.  At the same time, an HOD analysis of
observed galaxy clustering yields valuable tests of theoretical
models of galaxy formation.

The obtainable constraints on cosmological parameters depend, of course,
on the precision of the clustering measurements. 
Our assumption of 10\% fractional uncertainty on each observable
is probably conservative, and current galaxy clustering data have
already yielded higher precision for some of these observables,
such as the two-point correlation function.
To study the impact of improving measurement precision, we change the
assumed error bars of different sets of observables by turns 
and recalculate $\dchi2$ contours.  This investigation also highlights
the relative cosmological sensitivity of the different observables.
Each panel of Figure~\ref{fig:chgerr} (except for 
the top left panel discussed previously)
shows changes in the ($\Omega_m,\sigma_8$) constraints 
caused by improving the measurement precision of one set of 
observables. The two dotted contours are the 1-$\sigma$ and 2-$\sigma$ 
confidence levels ($\dchi2=2.30,$ 6.17) obtained by assuming
10\% fractional errors for all sets of observables; they match the 
solid contours in Figure~\ref{fig:contour}$a$. The solid contours are 
obtained by assuming 5\% errors for the set of observables indicated in 
the upper right corner of the panel, keeping 10\% for other sets.

With the galaxy bias factor constrained by other observables, a better 
measurement of galaxy density contrasts $\sigma_g$ leads to a better
constraint on $\sigma_8$, shrinking contours at both ends.
The tighter $\sigma_8$ constraint in turn produces a slightly tighter
$\Omega_m$ constraint, but there is essentially no improvement
in the $\sigma_8\Omega_m^q$ constraint (i.e., the width of the
contours in the narrow direction is unchanged).
Reducing the observational error bar on $Q_g$ produces
a similar improvement for a different reason --- the more precise 
measurements of $Q_g$ allow a 
better constraint on the galaxy bias factor $b_g$, so they help to break 
the degeneracy between $b_g$ and the amplitude of the dark matter power 
spectrum. 

Improving the pairwise velocity dispersion measurements produces modest
improvements in both the $\sigma_8\Omega_m^q$
constraint and the individual constraints on $\sigma_8$ and $\Omega_m$.
The former improvement comes mainly from the large-scale velocity
dispersion, which depends on halo velocities and thus scales
roughly with $\sigma_8\Omega_m^{0.6}$.  The individual parameter
improvements come mainly from the interplay of small-scale $\sigma_v$
with the apparent virial mass $\Mviravg$.  Better $\sigma_v$
measurements yield tighter constraints on the velocity bias
parameter, and the small-scale $\sigma_v$ and
$\Mviravg$ measurements then constrain the halo mass scale.
Improving the $\Mviravg$ measurements on their own yields better
$\sigma_8$ and $\Omega_m$ constraints but does not change the
$\sigma_8\Omega_m^q$
constraint.

Improving the $\xigg(r)$ measurements leads to a significantly
better determination of $\sigma_8$ for two reasons: it tightens the 
constraints on $P(N|M)$ and thus on the large-scale galaxy bias factor, 
and it improves the measurement of the large-scale galaxy clustering 
amplitude itself.  These two effects are analogous to those discussed 
earlier for $Q_g$ and $\sigma_g$, respectively.
Although $\xigg(r)$ is not directly sensitive to $\Omega_m$,
the tighter $\sigma_8$ constraint produces a tighter $\Omega_m$
constraint when combined with dynamical measurements.
Improving $\ngrp(\ge N)$ measurements produces surprisingly
little improvement in the cosmological parameter constraints.
If we instead degrade the fractional errors on $\ngrp(\ge N)$ to 50\%,
the constraints (shown by light solid contours) get noticeably
but not dramatically worse.
While it is always possible to reproduce $\ngrp(\ge N)$ exactly
by choosing the appropriate $\Navg$, this choice pins down
the mass associated with a given multiplicity, and it largely
pins down the galaxy bias factor.  
It appears, however, that the $\xigg(r)$ measurements already determine
the high end of $\Navg$ well enough for these purposes, 
so that moderate improvements in $\ngrp(\ge N)$ do not add
much power.  
If we degrade the presumed measurement errors on
$\xigg(r)$ to 25\%, then the impact of improving the $\ngrp(\ge N)$
measurements is larger, but it is still not dramatic.
The apparent redundancy 
of the $\ngrp(\ge N)$ and $\xigg(r)$ constraints makes them a valuable
consistency check on the underlying assumptions of the HOD modeling.
For example, the bias factor of single-galaxy halos affects
$\xigg(r)$ but not $\ngrp(\ge N)$, so an HOD model that makes the wrong
assumption about this bias factor will lead to conflicting
estimates of $\Navg$ at high masses.

We have also examined the consequences of adopting
priors on the HOD parameters, which could be motivated
by theoretical or observational considerations.
For example, numerical simulations give some idea of the plausible
range of values for the velocity bias factor,
and Tully-Fisher measurements can put limits on HOD parameters describing the 
cutoff profile of the mean occupation function. We characterize the priors by
a 1-$\sigma$ deviation allowed from a central value and minimize the sum
of the $\chi^2$ function and a quadratic term for the prior (see 
Appendix~\ref{sec:appendixB}). With a 1-$\sigma$ range of $1.00 \pm 0.15$
for the velocity bias factor $\alpha_v$ or $1.00 \pm 0.10$ for the 
distribution width parameter $\omega$, we find that there is almost no
improvement in cosmological constraints. The reason is that, even without
these priors, the measurements have already excluded cosmological models 
that need extreme values of $\alpha_v$ and $\omega$. For example, models 
that need $\alpha_v$ as high as 1.3 have $\Omega_m\lesssim 0.2$ and 
$\sigma_8 \lesssim 0.7$ (see Figures~\ref{fig:hodpar_omega} and 
\ref{fig:hodpar_sig8}). Adopting a prior for the halo concentration parameter 
$\Delta\log c_0$ with 1-$\sigma$  range of $0.0\pm0.2$ only slightly improves 
the cosmological constraints by excluding some small regions of extremely high 
or low values of $\sigma_8$. 
Priors on these HOD parameters must therefore be quite strong before they
would improve the constraints obtainable from the observable
set considered here.

As we have shown in Figure~\ref{fig:mcmc_hodpar}, for a fixed cosmology, 
the cutoff profile of the HOD cannot be well-constrained with the observables 
we adopt in this paper. This result
suggests that priors in the {\it shape} of the 
cutoff profile would not improve constraints on cosmological parameters. 
To verify this expectation, 
we perform a test in which we fix the cutoff profile width
$\sigM$ to the central model value.  The resulting
cosmological constraints are virtually identical to those in 
Figure~\ref{fig:contour}, where $\sigM$ is not fixed. However, if
the value of the absolute cutoff mass scale can also be pinned down,
e.g. from the Tully-Fisher relation or other dynamical measurements
in the single-galaxy regime, then cosmological constraints improve
substantially.
We have investigated the idealized case in which 
$\Mmin$ and $\sigM$ are fixed to the values of the central model.
We maintain the galaxy number density by
adjusting the amplitude of the satellite galaxy occupation function. 
The strong constraint on the halo mass scale from the value of $\Mmin$
produces a tight constraint on
$\Omega_m$ ($0.300\pm 0.008$), which in turn leads to a much
tighter constraint on $\sigma_8$ ($0.90\pm0.04$). 

Finally, we consider the constraints achievable with a brighter,
lower density galaxy sample.  Different 
samples of galaxies have different HODs, but they should probe the 
same underlying cosmology. All the investigations we have presented so 
far are based on the luminosity-threshold sample with a number density 
$\ngavg=0.01 h^3{\rm Mpc}^{-3}$, roughly corresponding to the
$M_r<-19.5+5\log h$ galaxies in the SDSS. We increase the mass scale of our 
central model HOD to construct a galaxy sample with number density 10 times 
smaller ($\ngavg=0.001 h^3{\rm Mpc}^{-3}$), which approximates the 
SDSS $M_r<-21+5\log h$ sample (\citealt{Zehavi05}). With the same 
assumption of 10\% fractional error on each of the 30 observables, 
we find that the cosmological constraints from these two galaxy samples
are quite similar, with the degeneracy direction 
$\sigma_8\Omega_m^q={\rm const.}$
slightly rotated to higher $q$ for the lower density sample. 
Thus, measurements for these two largely independent galaxy samples
should allow an important consistency check on cosmological conclusions
and, if the results are combined, a $\sim \sqrt{2}$ reduction in
statistical errors.

\subsection{The ``Influence Matrix''}
\label{sec:influence}

\begin{figure*}
\plotone{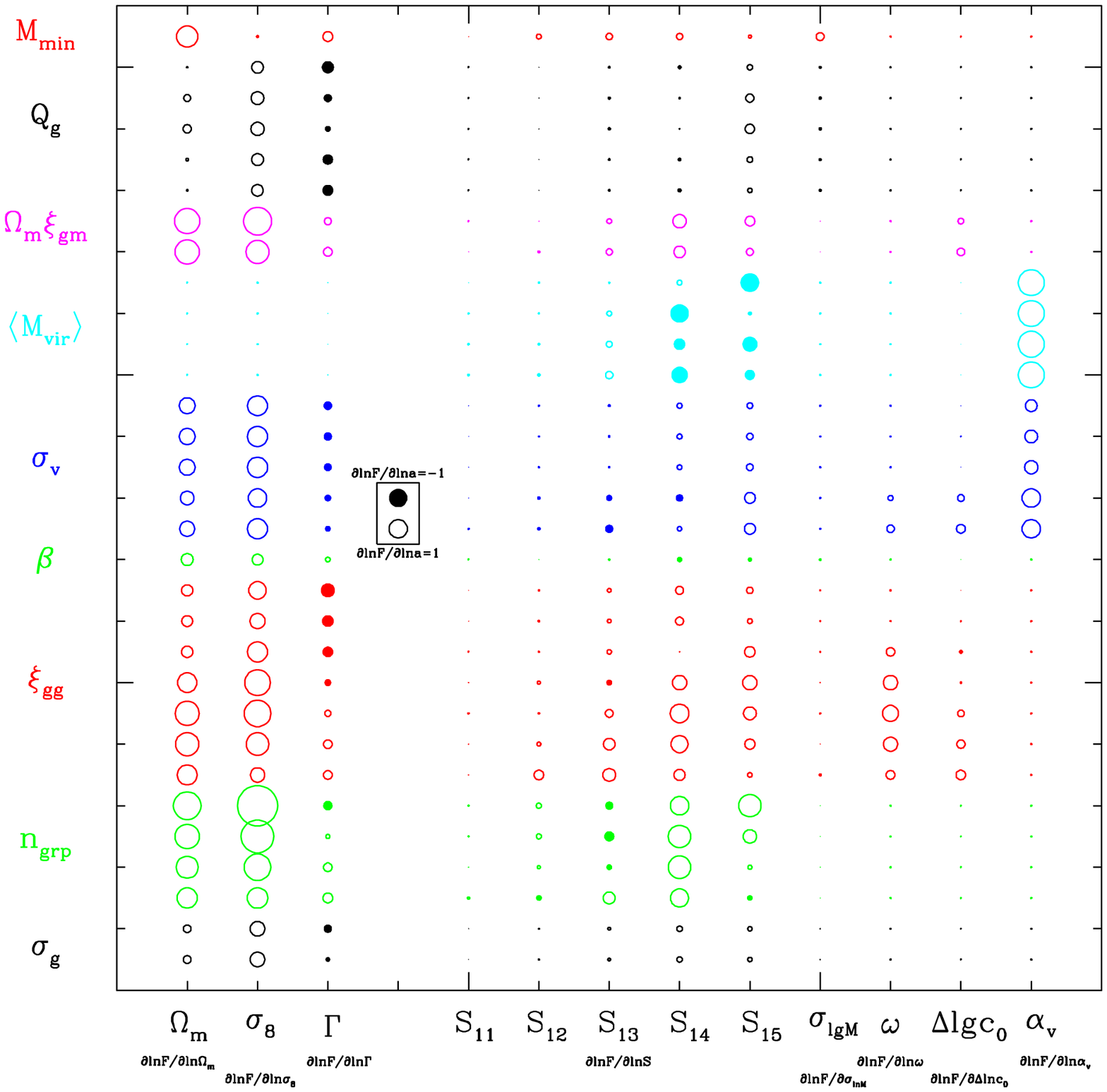}
\caption[]{\label{fig:matrix}
Logarithmic derivatives of the observables with respect to the cosmological
and HOD parameters, starting from the central model.
The area of each point is proportional to $\partial\ln F/\partial\ln a$,
where $F$ is the observable listed on the vertical axis and $a$ is
the parameter listed on the horizontal axis.  Open circles indicate
positive derivatives and filled circles negative derivatives, and
the scale is shown in the inset box.  For observables with multiple
scales, the smallest scale is shown at the bottom of the corresponding
set of points and the largest scale at the top, and color coding to match
circles to observables.
In our modeling, the HOD parameter $\Mmin$ is determined by matching the
galaxy number density once the other HOD parameters are set.
We show it here as an ``observable,'' whose value could in principle
be probed by dynamical or weak-lensing measurements.
}
\end{figure*}

Perhaps the most important implication of Figure~\ref{fig:chgerr}
is that the cosmological constraints do not rely on one or two
clustering statistics but instead emerge from the interlocking
web of measurements.  Figure~\ref{fig:matrix} demonstrates this
point in a different way.  It shows the ``influence matrix,''
which we define in terms of the partial derivatives
$\partial \ln F_i/\partial \ln a_j$ 
evaluated at the central HOD and cosmological model.
Each predicted observable 
$F_i$ is a function of cosmological and HOD parameters, and each $a_j$ 
represents one of these parameters, so the influence matrix encodes
the response of the observables to the individual model parameters.
When evaluating the partial derivatives, we adjust $\Mmin$ so that the 
galaxy number density stays fixed, but we keep the remaining HOD and
cosmological parameters (other than $a_j$) fixed.
The influence matrix is closely related to the Fisher matrix,
$$
{\partial^2 \ln{\cal L} \over \partial a_i \partial a_j} = 
\left({\partial {\mathbf F} \over \partial a_i}\right)^T {\mathbf C^{-1}} 
  \left({\partial {\mathbf F} \over \partial a_j}\right),
$$
where ${\mathbf F}$ is the vector of the predicted observables and
$\mathbf C$ is the covariance matrix of the errors. 
However, while the Fisher matrix depends strongly on the assumed error
covariance matrix $\mathbf C$, the influence matrix is independent of this
assumption, and it reveals the dependence of observables on model
parameters in a more transparent manner. 

Figure~\ref{fig:matrix} shows the influence matrix calculated at the 
central model. The area of each circle is proportional to the element
of the matrix from the corresponding observable ({\it vertical axis}) and
parameter ({\it horizontal axis}). The HOD parameter $\Mmin$, determined by
matching the galaxy number density, is presented here as an observable,
since its value could in principle be probed by dynamical or weak-lensing
measurements in the single-galaxy regime.
If two columns have similar elements, then the corresponding
parameters are largely degenerate, since an increase in one can be
compensated by a decrease in the other.

The first three columns of Figure~\ref{fig:matrix} show how clustering 
observables vary with cosmological parameters. 
Most elements of 
$\partial \ln {\mathbf F}/\partial \ln \Omega_m$ and
$\partial \ln {\mathbf F}/\partial \ln \sigma_8$ are positive.
An increase in $\Omega_m$ shifts the halo mass function to higher
mass scales (Fig.~\ref{fig:mf_xim}), and with the fixed satellite HOD,
this shift increases the group multiplicity function $\ngrp$. To keep a fixed
galaxy number density, $\Mmin$ increases by a larger factor than
$\Omega_m$, and the shift of galaxies to halos with larger $M/M_*$ leads
to stronger galaxy bias
for spatial clustering statistics $\sigma_g$ and $\xigg$.
An increase in $\sigma_8$ boosts the matter clustering directly.
Raising either parameter increases halo masses and velocities, and
thus the values of dynamical observables.
The similar form of these vectors implies significant degeneracy
between these two parameters, with an overall trend that
roughly follows 
$\partial \ln {\mathbf F}/\partial \ln \Omega_m = 0.75 
\partial \ln {\mathbf F}/\partial \ln \sigma_8$,
but the degeneracy is not complete.
The power spectrum shape parameter $\Gamma$ is largely decoupled
from $\Omega_m$ and $\sigma_8$.
The much stronger sensitivity of $\Mmin$ to $\Omega_m$ than to
any other cosmological or HOD parameter explains why a tight
observational constraint on $\Mmin$ can produce such a tight
constraint on $\Omega_m$ (\S\ref{sec:general}). The sensitivity of spatial 
clustering statistics to $\Omega_m$, which contrasts with the flat 
$\Delta\chi^2$ shown by the dotted line in Figure~\ref{fig:chi2_omega}, 
arises because we here keep all mass scales in the HOD fixed except for
$\Mmin$. If we instead scale all masses $\propto \Omega_m$, then the 
derivatives of $\xigg$, $\ngrp$, and $\sigma_8$ with respect to $\Omega_m$
essentially vanish, as expected, while the derivatives of $\Mviravg$
become one instead of zero.

The rightmost nine columns in Figure~\ref{fig:matrix} show how clustering
observables vary with the HOD parameters, with cosmological parameters
held fixed.
A degeneracy between a cosmological parameter and HOD parameters
exists to the extent that the ``influence vector''
$\partial \ln {\mathbf F}/\partial \ln a_C$ 
can be approximated by a linear combination of the influence vectors
$\partial \ln {\mathbf F}/\partial \ln a_{H,i}$,
where $a_C$ is the cosmological parameter and $a_{H,i}$ are
the HOD parameters.  As expected from our earlier results,
a substantial degeneracy exists if we restrict ${\mathbf F}$
to the spatial clustering observables $\xigg$, $\ngrp$, and $\sigma_g$.
For example, an increase in $\sigma_8$ can be largely compensated
by decreases in the parameters $S_{14}$ and $S_{15}$ that describe
the high end of $\Navg$, and changes in $\omega$ and $\dlgc0$ can
take up some amount of residual difference.  However, with the full set of
observables, the structures of 
$\partial \ln {\mathbf F}/\partial \ln \Omega_m$, 
$\partial \ln {\mathbf F}/\partial \ln \sigma_8$, and
$\partial \ln {\mathbf F}/\partial \ln \Gamma$ are
quite different from any of the influence vectors of
individual HOD parameters. 

\begin{figure*}
\plotone{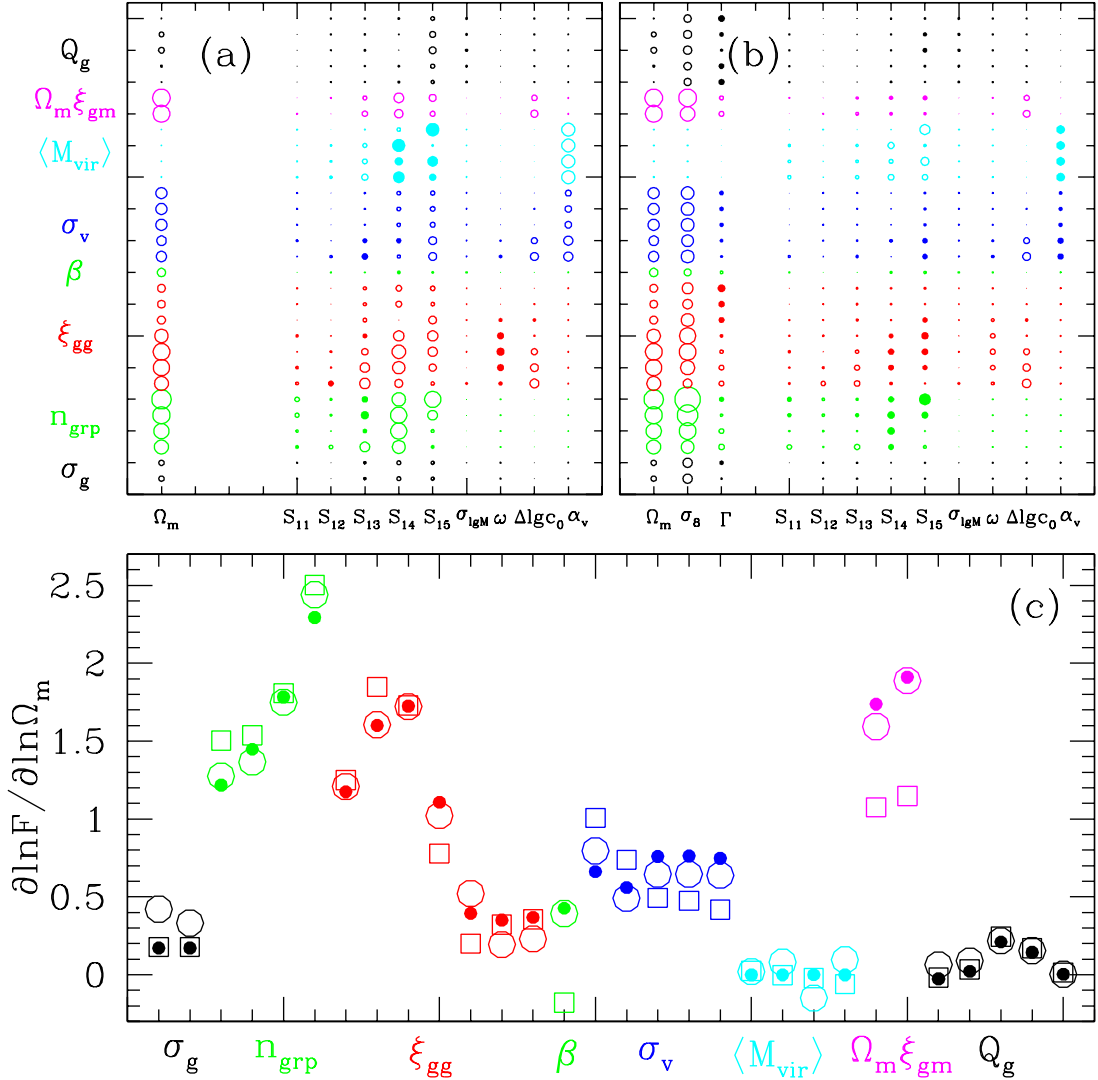}
\caption[]{\label{fig:lincomb}
Approximation of the $\Omega_m$ influence vector by a linear combination
of other influence vectors. Panels ($a$) and ($b$) are in a format similar to
Fig.~\ref{fig:matrix}, but each influence vector (each column) is multiplied
by the corresponding best-fit linear combination coefficient (except for the
$\Omega_m$ influence vector in the first column). Panel ($a$) shows the case
that keeps $\sigma_8$ and $\Gamma$ fixed and approximates the $\Omega_m$
influence vector by the HOD influence vectors only, and panel ($b$) shows
the case
that the $\Omega_m$ influence vector is approximated by the combination of
the $\sigma_8$, $\Gamma$, and HOD influence vectors. Panel ($c$) shows the
$\Omega_m$ influence vector ({\it filled circles}) and the best-fit linear
combinations excluding the $\sigma_8$ and $\Gamma$ influence vectors
({\it open squares}) and including them ({\it open circles}).
The open points track but do not completely overly the filled points,
which indicates partial but not complete degeneracy between $\Omega_m$
and the other model parameters with respect to these observables.
}
\end{figure*}

The visual structure of Figure~\ref{fig:matrix} suggests that there is no 
linear combination of the HOD influence vectors that accurately 
approximates the 
cosmological parameter influence vectors. We verify this suggestion
by solving for the 
combination coefficients that minimize the squared difference between 
the $\Omega_m$ influence vector and the corresponding
linear combination of the HOD influence vectors. 
(We do not include $\Mmin$ as an observable here, since it probably
cannot be measured as precisely as the other quantities.)
Figure~\ref{fig:lincomb}$a$ shows the HOD influence vectors multiplied by these
linear combination coefficients. The $S_{14}$ and $S_{15}$ parameters
dominate the best-fit linear combination, with $\alpha_v$ compensating for
the increase in $\Mviravg$ that would otherwise occur.
Figure~\ref{fig:lincomb}$c$ shows that this best-fit linear 
combination ({\it open squares}) does not fully recover the $\Omega_m$ 
influence vector ({\it filled circles}), with pairwise dispersions and 
(especially) galaxy-mass correlations providing the largest discrepancy.
If $\sigma_8$ and $\Gamma$ are allowed to vary in addition to the HOD
parameters, then the best-fit linear combination 
({\it open circles} in Fig.~\ref{fig:lincomb}$c$)
is a better approximation to the $\Omega_m$ influence vector,
with clear improvement in $\sigma_v$ and $\Omega_m\xigm$.
Figure~\ref{fig:lincomb}$b$ shows that 
the contribution from the $\sigma_8$ influence vector 
is much larger than that from any of the HOD influence vectors.
Nonetheless, the changes in $S_{14}$ and $S_{15}$ significantly improve
the agreement with the multiplicity function and small-scale correlation
function, relative to a pure $\sigma_8$ change with no freedom in the HOD.
As argued earlier,
the shape parameter $\Gamma$ is largely decoupled from the strong
cosmological degeneracy between $\Omega_m$ and $\sigma_8$.

Figure~\ref{fig:matrix} also shows that clustering observables are 
insensitive to some HOD parameters, most notably $S_{11}$ and 
$\sigma_{\rm logM}$, which describe the
low-mass end of the mean occupation function. 
We reached similar conclusions from our analysis of $\Navg$ 
constraints in \S\ref{sec:fixed} and the effect of a $\sigM$
prior in \S\ref{sec:general}.
The insensitivity 
of clustering observables to these parameters means that it is
difficult to constrain them with
clustering data. However, this insensitivity
also means that poor knowledge of these parameters does not introduce 
uncertainty in the cosmological conclusions. We have also investigated
the influence matrix for a simpler HOD parameterization in which the
mean satellite occupation function is a truncated power-law 
(\citealt{Kravtsov04,Zheng05}). The influence vector of the power-law
normalization is similar to that of $S_{14}$ in Figure~\ref{fig:matrix},
while the influence vector of the power-law slope is similar in overall 
form but with a different mass and length-scale dependence, reflecting its
differential impact on high- and low-mass halos. The influence of the 
low-mass cutoff is, again, weak. 

\section{Environmental Variation of the HOD}
\label{sec:environment}

The halo occupation function $\PNM$ is, by definition, a distribution
weighted by halo number, the probability that a randomly selected
halo of mass $M$ contains $N$ galaxies.  Our clustering calculations
implicitly assume that this distribution and other
parameters of the HOD like $\dlgc0$ and $\alpha_v$ are
independent of the larger scale environment.
A systematic dependence of a halo's galaxy occupation on 
the overdensity $\delta$ of its surroundings would alter the clustering
predicted for a given halo population and (number-averaged)
$\PNM$.  Ignoring this dependence could therefore lead one to
infer an incorrect HOD or incorrect cosmological parameters when
fitting observed galaxy clustering.

The assumption of an environment-independent HOD is rooted partly
in the \cite{Bond91} excursion set derivation of the extended
Press-Schechter formalism.  This derivation predicts that the
statistical features of a halo's assembly history depend only 
on its present mass, a point emphasized by \cite{White96}.
In agreement with this prediction, \cite{Lemson99} and
\cite{Sheth04} found that $N$-body halos of fixed mass in different
environments have similar properties and formation histories,
although \cite{Sheth04} found a subtle trend for close pairs of halos
to have higher formation redshifts.  The resolution of the 
simulations used in these studies effectively limited them
to halo masses $M \ga 10^{13}M_\odot$.  Two recent studies
\citep{Gao05,Harker06}, based on the higher resolution,
Millennium Run $N$-body simulation \citep{Springel05}, show that
the correlation between halo formation time and large-scale 
environment becomes much stronger for halo masses
$M \la 10^{12.5}M_\odot$.  To the extent that 
galaxy properties depend on halo formation time and
halo mass, this correlation will in turn produce an environmental
dependence of $\PNM$.

There are several reasons for thinking that the quantitative impact
of such a dependence is small, at least for galaxy samples defined
by simple thresholds in luminosity or baryonic mass.  Empirically,
\cite{Blanton06} show that the correlation of SDSS galaxy properties
with galaxy overdensity on a scale of $8\hmpc$ can be entirely
explained by the correlation with overdensity on the $1\hmpc$ scale
characteristic of large halos; at fixed local overdensity, they detect
no residual correlation with the larger scale environment.
On the theoretical side, \cite{Berlind03} show that the HOD
(more precisely, the mean occupation function) of galaxies above
a baryonic mass threshold in
\citeauthor{Weinberg04}'s (\citeyear{Weinberg04}) cosmological
hydrodynamic simulation is independent of environment within the
statistical uncertainties imposed by the $(50\hmpc)^3$ simulation
volume.  \cite{Yoo06}, analyzing the same simulation,
show that explicitly eliminating any environmental dependence by randomly
shuffling the galaxy populations among halos of similar mass changes
the galaxy-galaxy and galaxy-matter correlation functions by
$< 5\%$ on scales $0.1 \hmpc < r < 5\hmpc$, again within the statistical
uncertainties.  In a similar experiment using semi-analytic galaxy
populations in the Millennium Run simulation, \citet{Croton06}
find that the correlation functions of luminosity-bin samples
change by a few percent;
%(S. White, conference talk at U.C. Santa Cruz); 
because of the large simulation volume, these shifts are measured with
high precision.

Since we hope in the long term to achieve few percent precision on
cosmological parameter measurements, the Millennium Run results
imply that environmental dependence of the HOD must eventually
be considered as part of the program outlined in this paper.
Allowing arbitrary dependence of $\PNM$ on large-scale overdensity $\delta$
would probably make the problem intractable and degenerate, but
given the small magnitude of the anticipated effects, a more
restricted parameterization should suffice.  For example, since
the age dependence of halo clustering is found mainly in the mass
range of individual galaxy halos, it may well be enough to allow
$\Mmin$ to vary with $\delta$, or to treat the effective bias 
parameter of low-mass halos as a parameter to be fitted instead of using
the standard $N$-body result.  Since much of the signal
in most clustering statistics comes from multi-galaxy halos,
addition of such parameters may not have much impact on the
achievable constraints on cosmological parameters.
Semi-analytic mock catalogs will be valuable in guiding the
choice of parameterizations and in testing whether fitting
methods that adopt a particular parameterization (or that
ignore environmental variations altogether) yield incorrect
estimates of HOD or cosmological parameters.

For cosmological applications, it will be desirable to work with galaxy
samples defined by properties that are sensitive to halo
mass and relatively insensitive to formation history, such as
stellar mass or luminosity in a band dominated by old stellar populations.
Color-selected or morphologically selected samples
will be more susceptible to environmental dependence of halo formation
history at fixed mass.  
Conversely, samples with high luminosity or mass thresholds may be
the least susceptible, since the correlation of formation time with
environment is weak for high-mass halos.
While the breakdown of the \cite{Bond91}
prediction is an annoyance for the cosmological applications that
are the focus of this paper, it could be an asset for efforts to
understand the physical processes that determine galaxy morphology
and spectral properties.  For example, if fitting the clustering of a 
selected class of galaxies {\it requires} 
environmental dependence of the HOD, 
then the form and magnitude of that dependence will provide strong clues
to the aspects of halo formation that determine whether galaxies
are members of that class.  Conversely, if the color and
morphology dependence of observed clustering can be entirely 
explained without environmental variations of the HOD, it will
imply that aspects of halo formation that correlate with environment
do not play a strong role in determining these properties of galaxies.

\section{Summary and Discussion}
\label{sec:discussion}

Our analysis shows that HOD modeling can substantially increase the
cosmological power of galaxy clustering measurements, by breaking
degeneracies between the clustering of dark matter and the bias
of galaxies with respect to mass.  Changing the shape or amplitude
of the matter power spectrum or the value of $\Omega_m$ alters
the mass function, spatial clustering, and velocity statistics
of the dark halo population in well-understood ways
\citep{Zheng02}.  Our experiments here, 
which verify the qualitative arguments of
\cite{Berlind02} and \cite{Zheng02},
show that changes
to the galaxy HOD cannot mask these changes in the underlying
dark halo population. 
With our highly flexible
parameterization of the HOD, the set of observables considered
here yields $1\sigma$ uncertainties of $\sim 10\%$ in 
$\sigma_8$, $\Omega_m$, and $\Gamma$ and $\sim 5\%$ uncertainty
in the combination $\sigma_8 \Omega_m^q$ with $q \sim 0.75$.
We expect these forecasts to be conservative, as we have not
included observables for which we did not have ready analytic
approximations, and our assumption of 10\% measurement errors
is pessimistic in at least some cases.

The physical origin of these cosmological constraints is straightforward
to understand for simple changes in $\Omega_m$, $\sigma_8$, or
$\sigma_8 \Omega_m^{0.5}$, as discussed in 
\S\S\ref{sec:omegam}-\ref{sec:clnorm}.  
The general theme of these discussions is that, for a given
cosmological model, the spatial clustering
of galaxies largely determines the number of galaxies in 
halos of a given spatial abundance.
Dynamically sensitive statistics then reveal the halo mass scale,
which depends on $\sigma_8$ and $\Omega_m$.  We allow an arbitrary
bias $\alpha_v$ between the galaxy and dark matter velocity dispersions
within halos, but this freedom does not eliminate the constraining
power of dynamical observables because the space velocities of the
halos themselves do not change.  The main parameter degeneracy is
approximately $\sigma_8 \Omega_m^{0.75}$ because fixing this
combination roughly fixes the halo velocity scale and the abundance
of halos at the mass scale of rich galaxy groups.
However, the changing shape of the mass function, the differing
sensitivities of different velocity measures, and the different
$(\sigma_8,\Omega_m)$ dependence of galaxy-galaxy lensing all 
serve to break this degeneracy.  Figures~\ref{fig:chgerr} 
and~\ref{fig:matrix} demonstrate that the cosmological constraints
emerge from the full web of clustering observables and are
not dominated by one or two statistics on their own.

Constraints on the galaxy HOD will themselves provide valuable tests
of galaxy formation models.  The cutoff regime of $\Navg$ is 
difficult to pin down with the clustering statistics considered here,
but for a known cosmological model the relation between
average satellite number and halo mass is well determined,
and the satellite distribution width $\omega$, concentration $c_g$,
and velocity bias $\alpha_v$ are measured to 
$\sim 10\%$, 30\%, and 3\%, respectively (Fig.~\ref{fig:mcmc_hodpar}).
All of these quantities depend in detail on the physics that
governs the evolution of satellites in larger halos
(see, e.g., \citealt{Taylor04,Taylor05a,Taylor05b,Zentner05}), while the 
relative mass of halos that host central and satellite galaxies depends on 
the efficiency with which halos feed baryonic mass to their
central objects (see, e.g, discussions by \citealt{Berlind03,Zheng05}).
These galaxy formation constraints will be especially powerful
when derived as a function of luminosity, stellar mass, or other
observables.

The two main assumptions built into our modeling are the 
central-satellite parameterization and environment independence
of the HOD.  The central-satellite distinction appears well rooted
in galaxy formation physics, and it allows us to represent the
range of plausible galaxy HODs more completely with a moderate
number of parameters.  However, we have confirmed with other tests that
if we model galaxy bias with a flexible HOD parameterization that
does not impose a central-satellite distinction, but instead introduces
a characteristic mass for the narrow-to-wide transition of
$\PNM$ as in \cite{Scranton03}, then we reach almost identical
conclusions about the cosmological constraining power of the
clustering observables considered in this paper.  Recent numerical
results imply an environmental dependence of halo formation times
that opens the door to environmental variation of the HOD, especially
in the single-galaxy regime.  As discussed in \S\ref{sec:environment},
we expect the quantitative impact of such dependence to be small,
but potentially significant at the high level of precision we
ultimately hope to attain.  Investigation of environmental dependence
effects and methods of allowing for them in HOD modeling are
a high priority for future work.

The cosmological modeling approach advocated here is closely related
to the CLF method introduced by
\cite{Yang03} and \cite{Bosch03a}, who use clustering data and 
the global galaxy luminosity function to constrain the dependence
of the luminosity function on halo mass.
In principle, the CLF and HOD methods are equivalent --- they are
merely differential and integral forms of one another.
One can derive the CLF from a series of HOD fits to 
galaxy samples with different luminosity thresholds
\citep{Zehavi05,Tinker05}.  
Conversely, one can integrate the CLF to infer $\Navg$
for galaxies above a luminosity threshold \citep{Yang03,Bosch03b,Bosch05}.
The principal virtue of our HOD-based approach is that by
focusing on a single, well-defined class of galaxies, we can
parameterize the HOD in a way that seems likely to capture
the predictions of any reasonable galaxy formation model.
This kind of comprehensive parameterization is more difficult
to achieve for the full CLF, and most analyses to date have
assumed, for example, that the CLF has a Schechter form in halos
of fixed mass.  Nonetheless, it is valuable to pursue both HOD
and CLF approaches and test for consistency of conclusions.

HOD modeling complements rather than replaces the perturbative
approach based on large-scale measures that can be modeled with
linear or quadratic bias.  HOD modeling is more complex, but it
can take advantage of high-precision clustering measurements on small
and intermediate scales.  HOD modeling can also amplify the power
of the perturbative approach, extending its reach further into
the non-linear regime and checking its range of validity at a desired
level of precision.  For example, \cite{Tinker06} show that an
HOD-based approach to redshift-space distortions can improve
recovery of the perturbative parameter $\beta$ that controls
large-scale flows.  \cite{Yoo06} show that the linear bias model
provides an accurate description of galaxy-galaxy lensing for
$r \geq 2\hmpc$, and they show how to accurately model this
phenomenon on smaller scales (see also \citealt{Guzik02,Mandelbaum05}).
J. Yoo et al. (2007, in preparation) show that the scale-dependent bias
factors derived by fitting the projected galaxy correlation function
can extend recovery of the shape of the linear matter power
spectrum into the mildly non-linear regime.

CLF and HOD analyses of the 2dFGRS and SDSS redshift surveys
have already produced a number of interesting results, even
though they have considered only a fraction of the potential
clustering observables.  These results confirm, at least qualitatively,
many of the basic predictions of current galaxy formation models,
including the general form of the mean occupation function, the
dependence of this function on luminosity, the existence of a
minimum mass-to-light ratio in the halos of $\sim L_*$ galaxies
where galaxy formation is most efficient, the large gap between
the minimum halo mass for central and satellite galaxies above
a luminosity threshold, the sub-Poisson fluctuations of $\PNM$
that are a consequence of this gap, and the strong preference
of galaxies with older stellar populations for higher mass halos 
\citep{Bosch03a,Bosch03b,Magliocchetti03,Tinker05,Yang05,Zehavi04,Zehavi05,
Collister05}.
In combination with CMB data, the cosmological constraints from
HOD modeling of the SDSS projected galaxy correlation function
are almost as tight as those from the large-scale galaxy power
spectrum, and the two analyses are consistent within their
statistical uncertainties \citep{Abazajian05}.
For the most part, the cosmological inferences from CLF/HOD
modeling of galaxy clustering agree with those from other methods,
but matching the mass-to-light ratios of galaxy clusters simultaneously
with other clustering data appears to require values of $\Omega_m$
and/or $\sigma_8$ that are substantially lower than the commonly
adopted values of 0.3 and 0.9 \citep{Bosch03a,Tinker05}.
If this conclusion is correct, then the evidence for it should
become much stronger as more clustering observables are brought
into play and the SDSS data set itself moves to completion.\footnote{Since
the original submission of this paper, the analysis of the three-year 
{\it WMAP} data set \citep{Spergel06} has provided strong support for this 
shift in cosmological parameter values, in excellent agreement with the 
results of the HOD and CLF modeling and the related ``empirical model'' 
method of \citet{Vale06}.}

We have focused in this paper on the cosmological parameter constraints
that can be derived from galaxy clustering alone, using external
data only to guide the choice of power spectrum shape and motivate
the assumption of Gaussian initial conditions.  As with perturbative
analyses of large-scale structure, the long-term interest lies
in combining these constraints with those from the CMB,
the Ly$\alpha$ forest, Type Ia supernovae, and other cosmological
observables.  The complementary sensitivities of these observables
lead to much tighter parameter constraints.  More importantly,
conflicts among them could point the way to physics beyond the
simplest versions of $\Lambda$CDM, such as evolving dark energy,
a gravitational wave contribution to CMB anisotropy, departures from
scale invariance in the primordial power spectrum, non-zero space
curvature, cosmologically significant neutrino masses, and so forth.
By sharpening the constraints from large-scale structure in the new
generation of galaxy redshift surveys, HOD modeling can play a
critical role in efforts to test the standard cosmological model
and, perhaps, discover its breaking points.

\acknowledgments
We thank Andreas Berlind, Jeremy Tinker, and Jaiyul Yoo for
valuable discussions on these topics.
We thank Barth Netterfield for suggesting a cubic spline
parameterization of the mean occupation function, Sandy
Faber for suggesting the influence matrix investigation,
and Andy Gould for advice on the $\chi^2$ minimization method.
This work was supported by NSF grants AST 00-98584 and AST 04-07125. 
Z. Z. acknowledges the support of NASA through Hubble Fellowship grant 
HF-01181.01-A awarded by the Space Telescope Science Institute, which 
is operated by the Association of Universities for Research in Astronomy, 
Inc., for NASA, under contract NAS 5-26555. Z. Z. was also supported by a 
Presidential Fellowship from the Graduate School of the Ohio State University 
at an early stage of the project. 

\bigskip
\bigskip

%\centerline{\appendix{\bf APPENDIX}}
\appendix
\renewcommand{\thesection}{\Alph{section}}

\renewcommand{\theequation}{A\arabic{equation}}

\section{Pair Distributions of Two Populations Following NFW Profiles With
Different Concentrations}
\label{sec:appendixA}

For the calculation of the one-halo term of the galaxy-galaxy or galaxy-mass 
two-point correlation function, we need to know the distribution of galaxy 
pairs or galaxy-mass particle pairs as a function of separation in each halo.
The pair distribution is basically the convolution of the density profiles of 
two populations. For spherically symmetric density distributions $\rho_1$
and $\rho_2$, the fraction of pairs with separation in the range of $r$ to 
$r+dr$ is 
\begin{eqnarray}
\label{eqn:A0}
\frac{dF}{dr} dr & \propto &
 \int \rho_1({\bf x_1}) d^3 {\bf x_1} ~ \rho_2({\bf x_2+r}) d^3 {\bf r} \nonumber\\
                 & \propto &
 \int 4\pi x_1^2 \rho_1(x_1) dx_1 \int \rho_2(x_2) 2\pi r^2 \sin\alpha d\alpha dr \nonumber\\
                 & \propto &
 r^2 dr \int_0^{+\infty} dx_1 x_1^2 \rho_1(x_1) \int_{-1}^1 d\beta \rho_2(x_2),
\end{eqnarray}
where $F(r)$ is the fraction of pairs with separation less than $r$, 
$x_2^2=x_1^2+r^2-2x_1 r\beta$, $\beta=-\cos\alpha$, and $\alpha$ is the angle
between ${\bf x_1}$ and ${\bf r}$. If $\rho_1$ and $\rho_2$ follow the same
NFW profile truncated at the virial radius, the analytic expression of 
$dF/dr$ can be found in \citet{Sheth01a} (see their eq.[1] and eq.[A25]; 
note that
$\lambda(r)$ in their equations is proportional to $r^{-2}dF/dr$). In this 
paper we assume that, like the dark matter, the distribution of galaxies 
inside halos also follows the NFW profile. Therefore, we use the formula in
\citet{Sheth01a} when evaluating the one-halo term of galaxy-galaxy 
correlations. However, we 
allow galaxies to follow a different NFW profile than the dark matter in halos
of a given mass, which is our way to parameterize the spatial bias inside the 
halo in this paper. So, for the purpose of calculating the one-halo term of the 
galaxy-mass two-point correlation function,  we need the pair distribution 
function corresponding to two NFW profiles with different concentration 
parameters. 
We evaluate the convolution in equation~\ref{eqn:A0} for two NFW profiles with
concentration parameters $c_1$ and $c_2$, both truncated at the virial radius
$R_{\rm vir}$ of the halo. After tedious algebra, we derive the following
analytic expression for the (differential) pair distribution function 
$dF(x;c_1,c_2)/dx$, where $x\equiv r/(2R_{\rm vir})$.  

For $0\leq x\leq 0.5$ 
\begin{equation}
\label{eqn:A1}
\frac{dF(x;c_1,c_2)}{dx} = A(c_1,c_2) 
     [f_1(x;c_1,c_2)+f_2(x;c_1,c_2)+f_3(x;c_1,c_2)+f_4(x;c_1,c_2)]s,
\end{equation}
where
\begin{equation}
f_1(x;c_1,c_2) =  \frac{1}{(c_2+c_1+c_1 c_2 s)^2}\ln[(1+c_1 s)(1+c_2 s)]
                 +\frac{c_1 s}{c_2(c_2+c_1+c_1 c_2 s)(1+c_1 s)},
\end{equation}
\begin{equation}
f_2(x;c_1,c_2) =  \frac{1}{(c_2-c_1+c_1 c_2 s)^2}
                  \ln\left[\frac{(1+c_1 s)(1+c_2-c_2 s)}{1+c_1}\right]
                 - \frac{c_1(1-s)}{c_2 (c_2-c_1+c_1 c_2 s)(1+c_1 s)(1+c_1)},
\end{equation}
\begin{equation}
f_3(x;c_1,c_2) =  \frac{1}{(c_2-c_1-c_1 c_2 s)^2}
                  \ln\left[\frac{(1+c_2 s)(1+c_1-c_1 s)}{1+c_2}\right]
                 + \frac{c_1(1-s)}{c_2 (c_2-c_1-c_1 c_2 s)(1+c_1-c_1 s)},
\end{equation}
\begin{equation}
f_4(x;c_1,c_2) =  -\frac{s}{c_2(1+c_1)(1+c_2)(1+c_1-c_1 s)}, 
\end{equation}
and $s\equiv r/R_{\rm vir}=2x$. For the purpose of stableness, one should 
compute $f_2$ using the limit value $[c_1^{-2}-c_2^{-2}(1+c_1)^{-2}]/2$ as
$s$ approaches $(c_1-c_2)/(c_1 c_2)$ and similarly 
$f_3= [c_1^{-2}(1+c_2)^{-2} - c_2^{-2}]/2$ as $s$ approaches 
$(c_2-c_1)/(c_1 c_2)$. 

For $0.5< x\leq 1$,
\begin{equation}
\label{eqn:A2}
\frac{dF(x;c_1,c_2)}{dx} = A(c_1,c_2) f(x;c_1,c_2) s,
\end{equation}
where
\begin{equation}
f(x;c_1,c_2) = 
       \frac{1}{(c_2+c_1+c_1 c_2 s)^2}\ln
       \left[\frac{(1+c_1)(1+c_2)}{(1-c_1+c_1 s)(1-c_2+c_2 s)}\right]
      +\frac{s-2}{(1+c_1)(1+c_2)(c_2+c_1+c_1 c_2 s)}.
\end{equation} 

The distribution is normalized so that $\int_0^1 dx~ dF/dx=1$. 
The normalization factor $A(c_1,c_2)$ in equations~(\ref{eqn:A1}) and 
(\ref{eqn:A2}) can be expressed as
\begin{equation}
A(c_1,c_2)=\sqrt{A_*(c_1)A_*(c_2)},
\end{equation}
where we have the following fitting formula for $A_*(c)$,
\begin{equation}
A_*(c)=A_0 c^{3+\alpha} [1+(c/c_T)^{(\beta-\alpha)/\mu}]^\mu
\{1+B_0\sin [\omega(\log c-\phi)]\},
\end{equation}
with $A_0=2.4575$, $\alpha=-3.099$, $\beta=0.617$, $c_T=1.651$, $\mu=4.706$,
$B_0=0.0336$, $\omega=2.684$, and $\phi=0.4079$. The fitting formula has a 
fractional error less than 0.2\% for $1<c<100$.
 
It can be shown that the pair distribution derived here (eqs.[\ref{eqn:A1}]
and [\ref{eqn:A2}]) reduces to equation~(A25) in \citet{Sheth01a} for
the case $c_1=c_2$. We also test the above formulae by generating random 
particle distributions following different NFW profiles and counting particle 
pairs directly. The Monte Carlo result agrees with the analytic formulae 
perfectly.

\renewcommand{\theequation}{B\arabic{equation}}
\section{$\chi^2$ and Minimization}
\label{sec:appendixB}

We use the Gauss-Newton method to perform the $\chi^2$ minimization.
Here we briefly describe the main procedure. Readers are referred to 
\citet{Gould03} for more details.

Assume that we have $n$ observables $F_k$ ($k=1,2,...,n$) predicted 
by a model with $m$ parameters $a_i$ ($i=1,2,...,m$). 
Observations give $F_{k,{\rm obs}}$. The $\chi^2$ is defined as 
(Einstein's summation convention is used)
\begin{equation}
\chi^2 = [F_k({\bf a})-F_{k,{\rm obs}}]B_{kl}[F_l({\bf a})-F_{l,{\rm obs}}],
\end{equation}
where $B_{kl}$ is the inverse of the covariance matrix $\sigma_{kl}^2$ 
and has only diagonal components $B_{kk}=1/\sigma_{kk}^2$ if observational 
errors are uncorrelated.

If the initial guess of the solution of ${\bf a}$ is ${\bf a_*}$, through 
linearizing 
$F_k({\bf a})$ around ${\bf a_*}$ and demanding that the first derivatives
of $\chi^2$ are zero, we obtain an equation for the correction to the 
initial guess (\citealt{Gould03}),
\begin{equation}
 {\bf b \Delta a = d}, 
\end{equation}
where 
\begin{equation}
b_{ij} = \left. \frac{\partial F_k}{\partial a_i}\right|_* B_{kl} 
         \left. \frac{\partial F_l}{\partial a_j}\right|_*
       = \left. \frac{1}{2}\frac{\partial^2 \chi^2}{\partial a_i \partial a_j}
         \right|_*
\end{equation}
and
\begin{equation}
\label{eqn:d_i}
 d_{i} = \left. \frac{\partial F_k}{\partial a_i}\right|_* B_{kl} 
         [F_{l,{\rm obs}}-F_l({\bf a_*})]. 
\end{equation}
The direction of correction for ${\bf a_*}$ is given by ${\bf \Delta a}$. 
We set ${\bf a +} \epsilon {\bf \Delta a}$ as the new ${\bf a_*}$, 
where $\epsilon$ is a small number, and refine the estimation of the solution 
of ${\bf a}$.  The above process is repeated until $\chi^2$ reaches a minimum. 
It can be proved that the covariance matrix of ${\bf a}$ is just 
${\bf b^{-1}}$ (\citealt{Gould03}). 

If some parameters have additional constraints, we introduce a cost function
to realize the constraints and minimize the sum of $\chi^2$ and the cost 
function. For example, if parameter $a_1$ cannot exceed $a_2$, the cost 
function can be defined as $\mathcal{C}=[(a_1-a_2)/\sigma]^2$ for $a_1>a_2$ 
and $\mathcal{C}=0$ for $a_1\leq a_2$, where $\sigma$ is the tolerance. 
Another example is that a parameter has some priors. If $a_1$ is assigned
a prior of Gaussian distribution with mean $\hat a_1$ and standard 
deviation $\sigma$, the cost function can be defined as 
$\mathcal{C}=[(a_1-\hat a_1)/\sigma]^2$. Including a cost function leads to 
additional terms,  $-\partial \mathcal{C} / \partial a_i / 2$, in $d_i$ 
(eq.~[\ref{eqn:d_i}]).

\end{document}